\documentclass[a4paper,fleqn]{cas-dc}
\usepackage[numbers, sort&compress]{natbib}
\usepackage{hyperref}
\usepackage{multirow}
\usepackage{adjustbox}
\usepackage[flushleft]{threeparttable}
\hypersetup{hidelinks=true}
\usepackage{amsmath,amsfonts}
\usepackage{algorithmic}
\usepackage{array}
\usepackage[caption=false,font=normalsize,labelfont=sf,textfont=sf]{subfig}
\usepackage{textcomp}
\usepackage{stfloats}
\usepackage{url}
\usepackage{verbatim}
\usepackage{graphicx}

\def\tsc#1{\csdef{#1}{\textsc{\lowercase{#1}}\xspace}}
\tsc{WGM}
\tsc{QE}

\begin{document}
\let\WriteBookmarks\relax
\def\floatpagepagefraction{1}
\def\textpagefraction{.001}

\shorttitle{ELITE-RR}    

\shortauthors{}  

\title [mode = title]{A Skin-Tone-Aware Dual-Representation Remote Photoplethysmography Framework for Contactless Respiratory Rate Estimation}  

\tnotemark[1] 

\tnotetext[1]{ This work of Trishna Saikia is supported by the Prime Minister’s Research Fellowship (PMRF), the Ministry of Education, Government of India (2102743).} 

%

\author[1]{Trishna Saikia}
\ead{phd2201101014@iiti.ac.in}
\cormark[1]

\affiliation[1]{organization={Indian Institute of Technology (IIT) Indore},
            postcode={452020}, 
            country={India}}
\affiliation[2]{organization={University of Turku},
            postcode={20014}, 
            country={Finland}}

\author[1]{Anup Kumar Gupta}
\ead{msrphd2105101002@iiti.ac.in}

\author[1]{Puneet Gupta}
\ead{puneet@iiti.ac.in}

\author[2]{Pasi Liljeberg}
\ead{pakrli@utu.fi}

\cortext[1]{Corresponding author}

\begin{abstract}
\noindent\textbf{Background and Objective:} Respiratory rate is a vital indicator of pulmonary and cardiovascular health, yet conventional methods for estimating respiratory rate are often intrusive and unsuitable due to their contact-based nature. To this end, remote photoplethysmography offers a promising non-contact alternative and has been widely used for heart rate estimation; however, its potential for respiratory rate estimation remains underexplored. Existing methods typically adapt conventional remote photoplethysmography techniques, such as green and chrominance-based projections, originally designed for heart rate estimation, which only partially capture respiratory dynamics. Most prior work focuses on the Eulerian representation with fixed or empirically selected RGB projections, lacking adaptive mechanisms to extract maximal physiological information.

\noindent
\textbf{Methods:} To address these gaps, we propose a skin-tone-aware dynamic RGB signal projection that optimally captures respiratory information. To mitigate the sensitivity of the Lagrangian representation to non-respiratory motion, we introduce a dedicated denoising network for motion-based remote photoplethysmography signals. Furthermore, while some methods have initiated exploration of dual-representation fusion, they lack effective training strategies to transfer complementary information across representations. We address this limitation by designing a phase-independent contrastive loss that enables both representations to collaboratively learn relevant respiratory rate information. We also introduce RR-rPPG, a new respiratory-rate facial video dataset with Indian demographic representation, addressing the limited diversity of currently available resources.

\noindent
\textbf{Results:} We evaluate the proposed method on both the RR-rPPG dataset and the publicly available COHFACE dataset, where it consistently outperforms the comparison methods and achieves up to a 42.1\% reduction in mean absolute error across the evaluated settings.

\noindent
\textbf{Conclusion:} The proposed framework demonstrates the effectiveness of jointly leveraging skin-tone-aware Eulerian and denoised Lagrangian representations for contactless respiratory rate estimation from facial videos. In addition, the RR-rPPG dataset contributes a diverse benchmark resource for future research in remote respiratory monitoring. The code and dataset will be made publicly available upon paper acceptance.
\end{abstract}

\begin{graphicalabstract}
\includegraphics{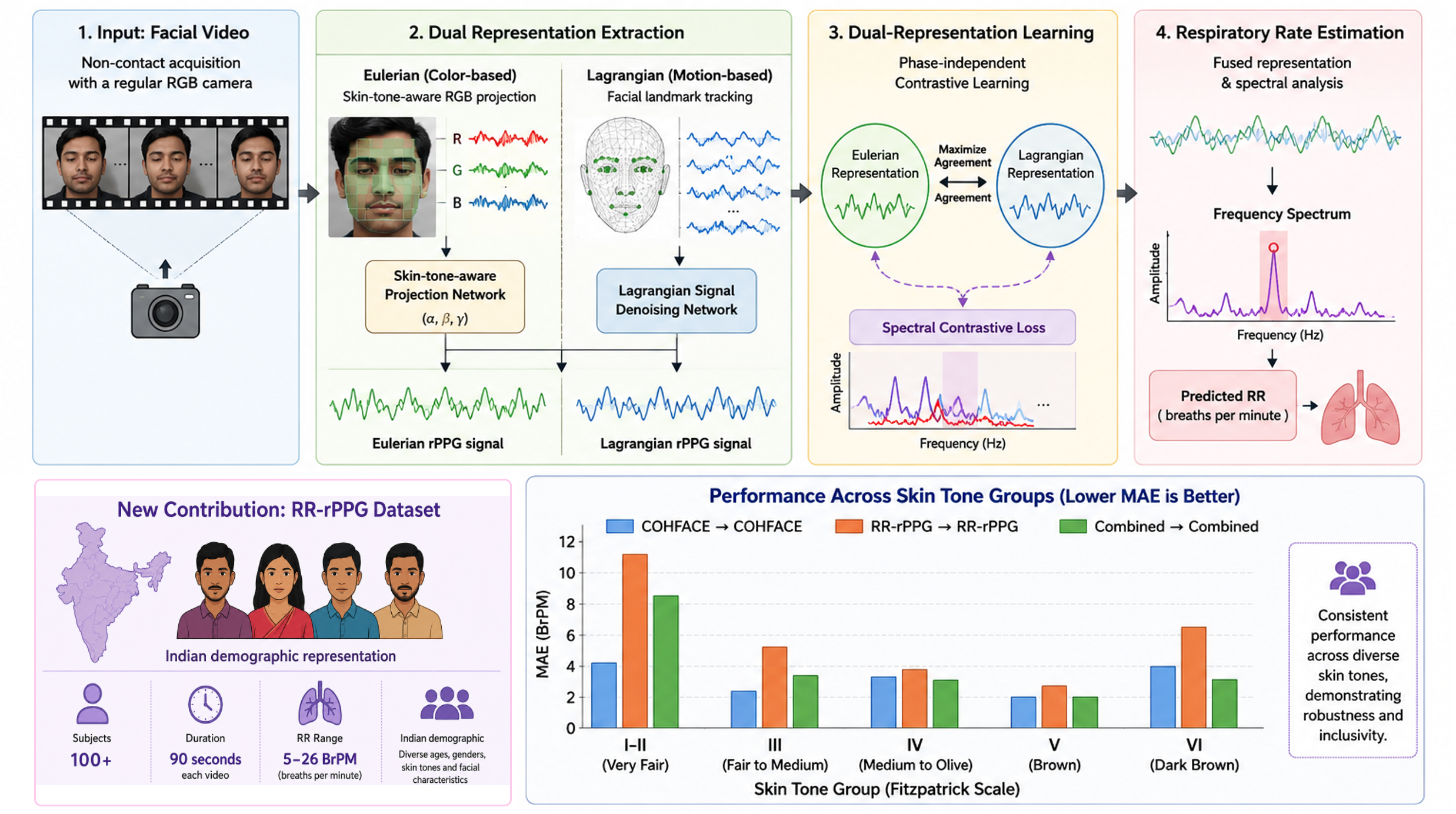}
\end{graphicalabstract}

\begin{highlights}
\item Dual-representation rPPG for contactless respiratory rate estimation.

\item Skin-tone-aware rPPG extraction improves respiratory signal quality.

\item RR-rPPG: An Indian demographic dataset for respiratory rate research.
\end{highlights}


\maketitle

\section{Introduction}
\label{sec:intro}
Respiratory rate (RR) is a vital physiological parameter that provides early insight into a person’s pulmonary and cardiovascular health. Abnormal RR patterns are often early indicators of critical conditions such as respiratory distress, cardiac arrest, and metabolic imbalances \citep{ryu2021measurement}. Conventional RR estimation methods, such as chest belts, spirometry, and capnography, require physical contact with the body, making them less suitable for continuous, unobtrusive, or remote health monitoring \citep{liao2025evaluation}. These constraints have driven significant interest in video-based physiological signal analysis methods, such as remote photoplethysmography (rPPG), which leverage regular RGB cameras to extract physiological signals \citep{gupta2022availability, gupta2023radiant}.

The rPPG method captures subtle skin color variations caused by blood volume changes in superficial facial vessels \citep{agarwal2025shine, lin2026facial}. These variations appear as temporal changes in red, green, and blue (RGB) pixel intensities across video frames \citep{birla2023ALPINE}. By processing these RGB signals, a composite physiological signal, referred to as the rPPG signal, can be derived \citep{du2021weakly}. While rPPG is primarily used for cardiac signal extraction, it can also reveal respiratory patterns due to respiratory sinus arrhythmia (RSA), a physiological phenomenon where heart rate (HR) varies in sync with the breathing cycle \citep{luguern2020assessment}. However, extracting clean respiratory signals from rPPG in real-world scenarios remains challenging due to factors such as skin tone diversity and other appearance-related variations.

In the context of rPPG-based RR estimation, two complementary visual modalities are employed: Eulerian and Lagrangian, which correspond to color-based and motion-based methods, respectively \citep{gwak2024multimodal}. Eulerian-based methods extract rPPG signals by analyzing subtle temporal variations in skin color intensity, typically in exposed facial regions \citep{liao2025evaluation}. 
In contrast, Lagrangian-based methods estimate RR by tracking periodic respiratory-induced displacements in facial regions, such as those on the nostrils or cheeks \citep{gwak2024multimodal}. Notably, Lagrangian-based methods do not require direct visibility of skin tissue and can remain effective even when portions of the face are occluded \citep{saikia2023hreadai}. However, they are more susceptible to confounding artifacts from facial expressions, speech, or head movements.

While Eulerian-based methods are comparatively robust to such motion artifacts, their performance is heavily influenced by skin tone diversity. This highlights the need for a dynamic, skin-aware projection algorithm that can adaptively extract high-fidelity rPPG signals from RGB signals based on skin characteristics. Additionally, the incorporation of a dedicated denoising network can further refine Lagrangian signals by suppressing non-respiratory noise. Finally, integrating both Eulerian and Lagrangian representations through a unified dual-representation framework holds the potential to improve the robustness of RR estimation across varied subjects.

Despite the complementary nature of Eulerian and Lagrangian methods, most rPPG-based RR estimation methods are predominantly built upon Eulerian methods such as methods in \citep{du2021weakly, luguern2020assessment, vatanparvar2022respiration}. To this end, many of these methods utilize fixed empirical projection algorithms like CHROM \citep{de2013robust} to extract respiration-induced rPPG signals from the RGB channels. However, such static methods overlook individual variations in skin tone, thereby limiting their generalizability and physiological fidelity. Moreover, the majority of existing methods treat Eulerian and Lagrangian signals independently \citep{van2016robust, du2021weakly}, failing to leverage their complementary strengths. Although some recent efforts have explored dual-representation integration for robust RR estimation \citep{gwak2024multimodal}, they generally lack an adaptive fusion mechanism that dynamically adjusts based on skin characteristics and signal quality. Furthermore, none of these methods explicitly facilitates cross-modal knowledge sharing to refine one representation using insights from the other. In response, we propose a unified, skin-aware method that not only combines both representations but also employs a novel phase-independent contrastive learning strategy. This enables alignment in the frequency domain across representation, despite temporal mismatches, thereby supporting joint learning and robust RR estimation.

Another limiting factor in advancing rPPG-based RR estimation is the scarcity of diverse, publicly available datasets specifically curated for this task.  Existing datasets are either limited in demographic diversity or not specifically curated for RR estimation task. Notably, there is an absence of datasets representing Indian skin tones and facial features, which restricts the applicability and generalizability of current methods across different population groups.

To address these limitations, we introduce RR-rPPG, a novel dataset comprising facial videos of subjects from diverse Indian demographics, accompanied by synchronized respiration and pulse signals. In addition to this, we propose \textit{ELITE-RR} (\textbf{\textit{E}}ulerian and \textbf{\textit{L}}agrangian \textbf{\textit{I}}ntegration with skin-\textbf{\textit{T}}one awar\textbf{\textit{E}}ness for \textbf{\textit{RR}} estimation), a dual-representation rPPG-based RR estimation method that adaptively combines Eulerian and Lagrangian methods through a quality-guided skin-aware fusion strategy. Our method is designed to enhance robustness across varying skin tones and improve physiological signal fidelity. Its key contributions are as follows:

\begin{enumerate}
    \item We propose a novel, data-driven method that adaptively transforms RGB pixel intensities into a physiological rPPG signal using skin tone-aware features. Unlike conventional methods that apply static or empirically derived projections, our method learns dynamic projection weights: $\alpha$, $\beta$, and $\gamma$, through a shallow neural network conditioned on handcrafted skin-tone-aware features. This enables personalized and robust signal extraction across varying skin tones.

    \item To harness the complementary information of Eulerian and Lagrangian representations, we propose a contrastive loss that aligns signals based on their frequency content rather than phase. This design accounts for phase misalignment across representations and allows our networks to learn robust, modality-agnostic representations focused on respiratory frequency content.
    
    \item We present the first rPPG dataset designed for RR estimation that includes Indian demographic diversity. This dataset will enable the development and benchmarking of more inclusive and generalizable methods in the field.
\end{enumerate}

\section{Preliminaries}
\label{sec:prelim}

\subsection{Remote Photoplethysmography (rPPG)}
The rPPG is a non-contact method that estimates physiological parameters like HR and RR by analyzing subtle changes in skin color captured through RGB cameras \citep{zhu2024comparative}. These color variations arise from blood volume fluctuations beneath the skin, which affect the amount of reflected light and appear as temporal changes in the RGB channels of facial videos \citep{chen2019respiratory}. In the rPPG domain, two complementary representations are commonly used: the Eulerian representation, which analyzes pixel intensity changes at fixed locations to detect color-based signals \citep{birla2023ALPINE}, and the Lagrangian representation, which tracks the motion of discriminative facial landmarks over time to capture movement-based signals \citep{gwak2024multimodal}. Respiration affects both representations: it influences HR through RSA, which alters the color-based signals, and it causes subtle head and facial movements that are captured in motion-based signals \citep{du2021weakly}.

\noindent
\subsection{RR Estimation Using Eulerian Representation:}
To extract respiratory signals using the Eulerian representation, earlier methods have employed signal decomposition techniques such as Independent Component Analysis (ICA) and Principal Component Analysis (PCA), or signal amplification approaches like Eulerian Video Magnification (EVM) \citep{luguern2020assessment}. In recent literature, methods such as \citep{vatanparvar2022respiration} and \citep{du2021weakly} primarily adopt the CHROM method to obtain the Eulerian rPPG signal. These methods typically rely on predefined rules or empirically derived projections to combine RGB channels. While such strategies perform well in controlled environments, they often fail to generalize across individuals with diverse skin tones. This makes the Eulerian modality particularly sensitive to appearance-related variations.

\subsection{RR Estimation Using Lagrangian Representation:}
Existing methods employing the Lagrangian representation estimate RR by extracting rPPG signals through the tracking of subtle facial motions, particularly around regions like the nostrils or cheeks \citep{gwak2024multimodal}. This modality has proven effective, especially in scenarios where skin visibility is limited, and has been shown to outperform Eulerian-based methods under such conditions \citep{van2016robust, saikia2023hreadai}. However, these motion signals are highly vulnerable to contamination from non-respiratory movements like head shifts or facial expressions. Despite this challenge, prior works have not explicitly addressed the denoising of such signals. To overcome these limitations, we design a dedicated denoising network aimed at refining the Lagrangian rPPG signals for robust RR estimation.

\subsection{Dual-Representation RR Estimation:}
Recent research has focused on dual-representation methods that fuse both Eulerian and Lagrangian representations to overcome the limitations of using either representation alone. This integration leverages both of their complementary strengths. A few existing methods estimate RR independently from each representation and then rely on a machine learning model to select the more reliable one \citep{gwak2024multimodal}. 
However, such methods do not enhance the performance of the complementary representation. They treat the two signal representations independently, lacking a mechanism for mutual refinement or knowledge transfer. To address these limitations, we design a unified dual-representation method that performs three coordinated tasks: (1) a projection network that dynamically fuses RGB signals based on skin tone variations, (2) a denoising network that enhances the reliability of Lagrangian signals, and (3) a novel training strategy using a phase-independent contrastive loss that enables the model to align spectral characteristics of signals across representations, despite possible phase differences, thereby facilitating effective knowledge transfer and joint learning for robust RR estimation. Moreover, the absence of demographically diverse datasets, particularly those representing Indian skin tones, restricts the generalizability of existing methods. In response, we introduce a new dataset specifically curated for this purpose.

\section{Proposed Method}
\label{sec:proposed_method}
\begin{figure*}
    \centering
\includegraphics[width=0.95\linewidth]{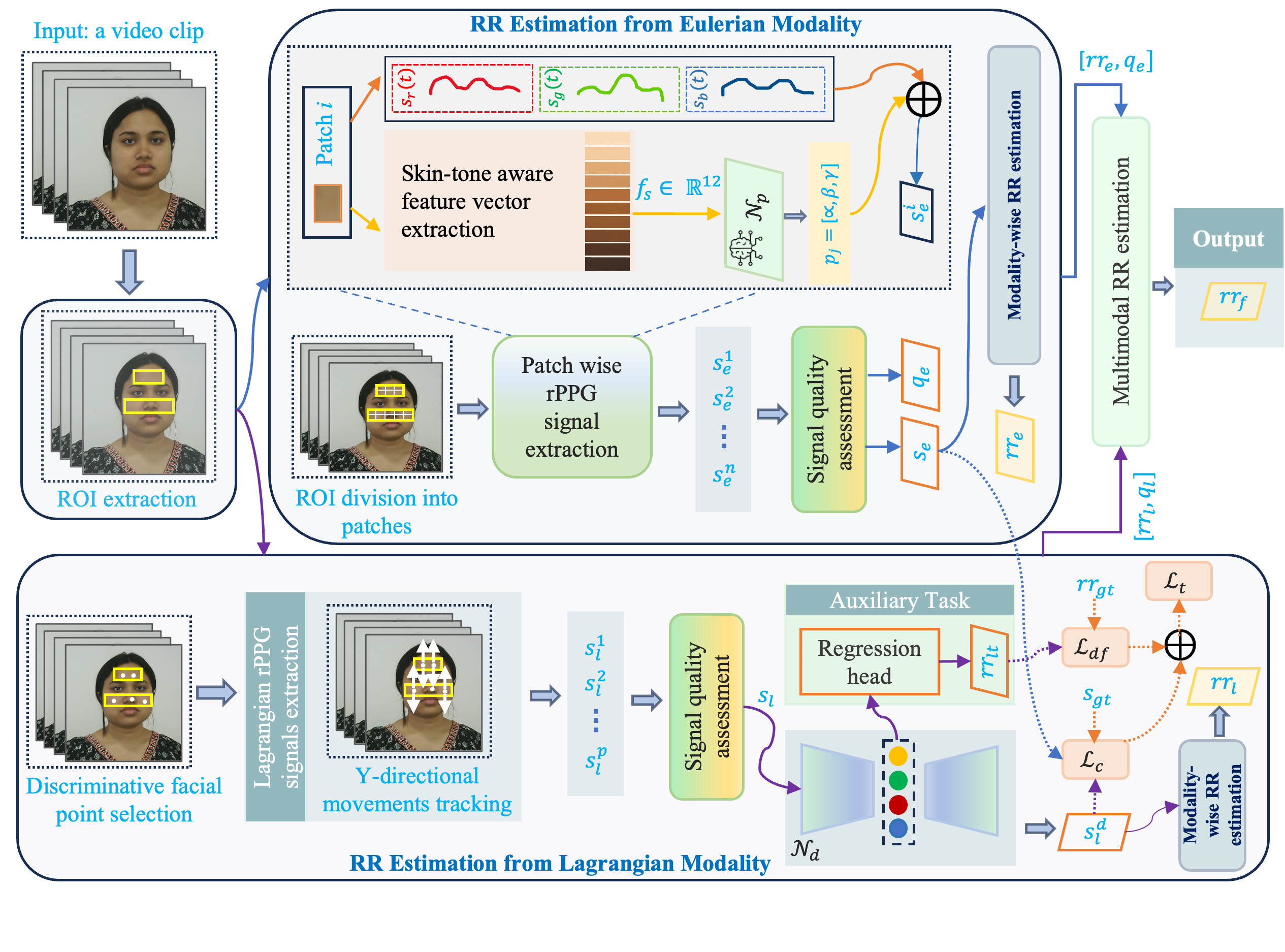}
    \caption{Overview of the proposed method, \textit{ELITE-RR}. It combines two complementary representations: Eulerian and Lagrangian. In the Eulerian branch (top), the facial ROI is divided into patches. Each patch's skin-tone-aware features guide a projection network ($\mathcal{N}_p$) to transform RGB signals into reliable physiological signals. In the Lagrangian branch (bottom), vertical displacements of facial landmarks are tracked to extract rPPG signals, which are refined using a denoising network ($\mathcal{N}_d$). A phase-independent contrastive loss ($\mathcal{L}_c$) facilitates knowledge transfer between representations. Finally, RR is estimated through a quality-weighted fusion of both representations.}\label{fig:work-flow}
\end{figure*}
We propose a dual-representation method, \textit{ELITE-RR}, for RR estimation from facial videos by integrating both Eulerian and Lagrangian representations (refer to Figure \ref{fig:work-flow}). The input video is divided into fixed-length non-overlapping clips, and for each clip, facial landmarks are detected to extract two regions of interest (ROIs): the forehead and cheeks (including nostrils). In the Eulerian branch, each ROI is divided into equal-sized patches. For every patch, we extract temporal RGB signals and compute a feature vector that encodes skin tone characteristics. A shallow neural network then predicts a set of dynamic projection weights to combine RGB channels into a single rPPG signal. The patch with the highest signal-to-noise ratio (SNR) is selected, and RR is estimated via spectral analysis. In the Lagrangian branch, vertical motions within the ROIs are tracked to capture respiration-induced movements. The trajectory with the best SNR is further denoised using a CNN-based network, and RR is computed from its frequency spectrum. Finally, RR estimates from both representations are fused using an SNR-based quality-weighted strategy, yielding a robust and enhanced RR estimation.

\subsection{ROI Extraction}
\label{ROI-extraction}
 We extract facial ROIs relevant for rPPG signal estimation from each video clip. At first, we utilize the MediaPipe Framework \cite{lugaresi2019mediapipe} to detect the face and its corresponding landmarks in each frame. These landmarks enable precise localization of facial sub-regions that are physiologically informative and less susceptible to expression-induced artifacts. We intentionally exclude areas around the eyes and mouth to avoid the influence of facial expressions, which can introduce noise into the rPPG signal. Instead, we focus on two primary regions that have been shown to contain strong respiratory-related signals: the forehead and cheeks (including the nostrils). The forehead ROI is defined by applying a minimum enclosing rectangle around the detected forehead landmarks as suggested in \citep{lokendra2022and}. Similarly, the cheek and nostril regions are extracted based on their respective landmarks. To further enhance signal quality and account for inconsistencies in rPPG information across different facial regions, each ROI is subdivided into smaller, non-overlapping square patches, as described in \citep{saikia2023hreadai}. This division not only facilitates localized rPPG signal extraction but also increases the robustness of signal processing by allowing the selection of high-quality patches during inference.

\subsection{RR Estimation from Eulerian Representation}
\label{eulerian-modality}
After extracting the facial ROIs and dividing them into smaller patches, we compute temporal signals of the red, green, and blue channels, denoted as $\mathbf{s}_r(t)$, $\mathbf{s}_g(t)$, and $\mathbf{s}_b(t)$, respectively. Each of these signals is obtained by averaging the corresponding color channel pixel values within the patch across video frames. To suppress noise and retain only respiration-relevant components, each signal is passed through a 4th-order Butterworth band-pass filter with a cutoff frequency of [0.08, 0.5] Hz, a range corresponding to normal respiratory frequencies \citep{vatanparvar2022respiration}. Finally, to project these filtered signals into a single physiological rPPG signal that maximally captures respiratory activity, we employ a shallow neural network, $\mathcal{N}_p$ that generates a dynamic projection vector $\mathbf{p}_j = [\alpha, \beta, \gamma]$ based on the patch-specific skin tone feature vector, $\mathbf{f}_s$. Details of the $\mathcal{N}_p$ architecture are provided in Section \ref{network-architecture}, while the construction of the skin-tone feature vector $\mathbf{f}_s$ is described in the next section.

\subsection{Skin-tone-aware Feature Vector}
\label{skin-tone-vector}
To adaptively extract high-quality rPPG signals from diverse facial regions, it is essential to account for variations in skin tone. For this purpose, we construct a 12-dimensional feature vector $\mathbf{f}_s \in \mathbb{R}^{12}$ for each patch, summarizing its color characteristics and temporal stability. This vector is used as input to a shallow neural network $\mathcal{N}_p$ (detailed in Section \ref{network-architecture}) to derive an optimal projection for rPPG signal extraction. The components of $\mathbf{f}_s$ are organized into four groups, each capturing a distinct visual property:

\subsubsection{Mean Channel Intensities} 
These values reflect the average color composition of the patch and provide a baseline estimate of the subject’s skin tone. For that, we calculate the mean RGB pixel intensities from the first frame of the video clip within the selected patch:
\begin{equation*}
\mu_R = \text{mean}(R_1), \quad
\mu_G = \text{mean}(G_1), \quad
\mu_B = \text{mean}(B_1)
\end{equation*}
Here, $R_1$, $G_1$, and $B_1$ denote the red, green, and blue pixel values of the selected patch in the first frame, respectively.

\subsubsection{Temporal Standard Deviations}
These features measure the temporal fluctuations in average channel intensities, capturing how stable the patch’s color is over time. Thereby reflecting appearance variability within the patch. It is calculated as follows:
\begin{align*}
\sigma_R &= \text{std}\left( \{\text{mean}(R_t)\}_{t=1}^{T} \right), \nonumber \\
\sigma_G &= \text{std}\left( \{\text{mean}(G_t)\}_{t=1}^{T} \right), \nonumber \\
\sigma_B &= \text{std}\left( \{\text{mean}(B_t)\}_{t=1}^{T} \right)
\end{align*}
Here, $std(\cdot)$ refers to the standard deviation operation.

\subsubsection{Chromatic Ratios}
Further, we compute inter-channel mean intensity ratios to capture the color balance of the patch as follows:
\begin{equation*}
\text{ratio}_{RG} = \frac{\mu_R}{\mu_G + \varepsilon}, \;
\text{ratio}_{RB} = \frac{\mu_R}{\mu_B + \varepsilon}, \;
\text{ratio}_{GB} = \frac{\mu_G}{\mu_B + \varepsilon}
\end{equation*}
These ratios capture color dominance, aiding in distinguishing skin tones variations.

\subsubsection{Normalized RGB Vector}
To isolate pure color composition from brightness, we normalize the mean RGB values of a patch to form a unit vector.

\begin{equation*}
\hat{\boldsymbol{\mu}}_{RGB}
=
\frac{1}{\sqrt{\mu_R^2+\mu_G^2+\mu_B^2+\varepsilon}}
\left[\mu_R,\ \mu_G,\ \mu_B\right]
\end{equation*}
This normalization reduces the influence of overall intensity changes while preserving the relative color composition of the patch.

\noindent
\textbf{Final Feature Vector:}
All computed features are concatenated to form the final skin-tone-aware feature vector $\mathbf{f}_s$, which can be represented as follows:
\begin{equation*}
    \begin{aligned}
    \mathbf{f}_s
    = [&\mu_R, \mu_G, \mu_B,\ \sigma_R, \sigma_G, \sigma_B, \\
    &\operatorname{ratio}_{RG}, \operatorname{ratio}_{RB}, \operatorname{ratio}_{GB},
    \hat{\mu}_R, \hat{\mu}_G, \hat{\mu}_B]
    \in \mathbb{R}^{12}
    \end{aligned}
\end{equation*}
The resulting feature vector provides a comprehensive summary of the patch’s skin tone properties. It is then fed into the projection network $\mathcal{N}_p$, which outputs a dynamic projection vector $\mathbf{p}_j = [\alpha, \beta, \gamma]$ used to compute the patch-level rPPG signal, $\mathbf{s}_e$ for the Eulerian modality:
$\mathbf{s}_e = \alpha \cdot \mathbf{s}_r(t) + \beta \cdot \mathbf{s}_g(t) + \gamma \cdot \mathbf{s}_b(t)$.
This process is repeated for all patches. In training, each patch is treated as a separate sample. At inference, we compute SNR for each patch’s signal (following \citep{saikia2023hreadai}) and select the one with highest SNR. Final RR is estimated using method in Section \ref{rr-estimation}.

\subsection{RR Estimation from Lagrangian Representation} 
After extracting the facial ROIs, we apply the Lucas-Kanade tracker \citep{balakrishnan2013detecting} to identify and track a set of discriminatory feature points across frames. For each point, we obtain a temporal trajectory that records its motion in both horizontal and vertical directions. Since vertical motion has been shown to be more sensitive to respiration-induced variations \citep{molinaro2022contactless}, we retain only the $y$-axis component of each trajectory. Let $n$ denote the total number of frames and $p$ the number of tracked points. The Lagrangian rPPG signals are represented as:
$\mathbf{S}_l = [\mathbf{s}_l^1, \mathbf{s}_l^2, \ldots, \mathbf{s}_l^p]$, where $\mathbf{s}_l^i = [v_1^i, v_2^i, \ldots, v_n^i]$ denotes the vertical motion trajectory of the $i^{\text{th}}$ point across $n$ frames.
Although these extracted signals reflect periodic respiratory-induced movements, they may still be contaminated by noise due to non-respiratory facial motion or tracking inaccuracies. To suppress irrelevant frequency components and retain only respiration-related variations, each signal is first passed through a band-pass filter with a frequency range of [0.08, 0.5] Hz, consistent with the typical respiratory frequency range \citep{vatanparvar2022respiration}. Following filtering, we evaluate the signal quality using the SNR metric and select the best-quality signal. This selected signal is then refined using a dedicated denoising network, $N_d$, which mitigates residual artifacts and enhances respiratory features. Finally, the denoised signal is subjected to spectral analysis to estimate the RR, following the procedure outlined in Section \ref{rr-estimation}.

\subsection{Representation-wise RR Estimation}
\label{rr-estimation}
Once the rPPG signals are obtained from each representation, RR is estimated by analyzing their frequency spectra using the Fast Fourier Transform (FFT), as outlined in \citep{gupta2017serial}. This is based on the observation that respiratory-induced oscillations in rPPG signals typically occur within a specific frequency range, while other components are attributed to unrelated physiological signals. The frequency corresponding to the highest spectral peak is assumed to represent the respiratory rate. Let $\mathbf{s_f}$ denote the FFT spectrum of the rPPG signal. The respiratory rate $rr$ (in breathing rate per minute, BrPM) is calculated as:
$rr =  m(\mathbf{s_f}) \times 60$. Here, $m(\mathbf{s_f})$ is the peak frequency (Hz); multiplying by 60 converts it to BrPM.

\subsection{Dual-representation RR Estimation}
\label{multi-rr-estimation}
While both Eulerian and Lagrangian representations independently provide valuable insights into respiratory activity, each has its own limitations. The Eulerian representation is sensitive to skin color variations, whereas the Lagrangian representation is prone to motion artifacts and tracking noise. To mitigate these limitations and improve robustness, we performed a fusion of the RR estimates from both representations. Specifically, we adopt a quality-weighted fusion strategy based on the SNR of the respective rPPG signals. Let $rr_e$ and $rr_l$ denote the respiratory rate estimates from the Eulerian and Lagrangian representation, and let $q_e$ and $q_l$ represent their corresponding signal qualities. The final fused RR estimate, $rr_f$, is computed as a weighted average: $rr_f = \frac{q_e \cdot rr_e + q_l \cdot rr_l}{q_e + q_l}$. This formulation ensures that the representation providing a cleaner signal (higher SNR) contributes more significantly to the final estimation. 

\subsection{Network Architecture Details}
\label{network-architecture}

\begin{figure}
    \centering
    \includegraphics[width=0.98\linewidth]{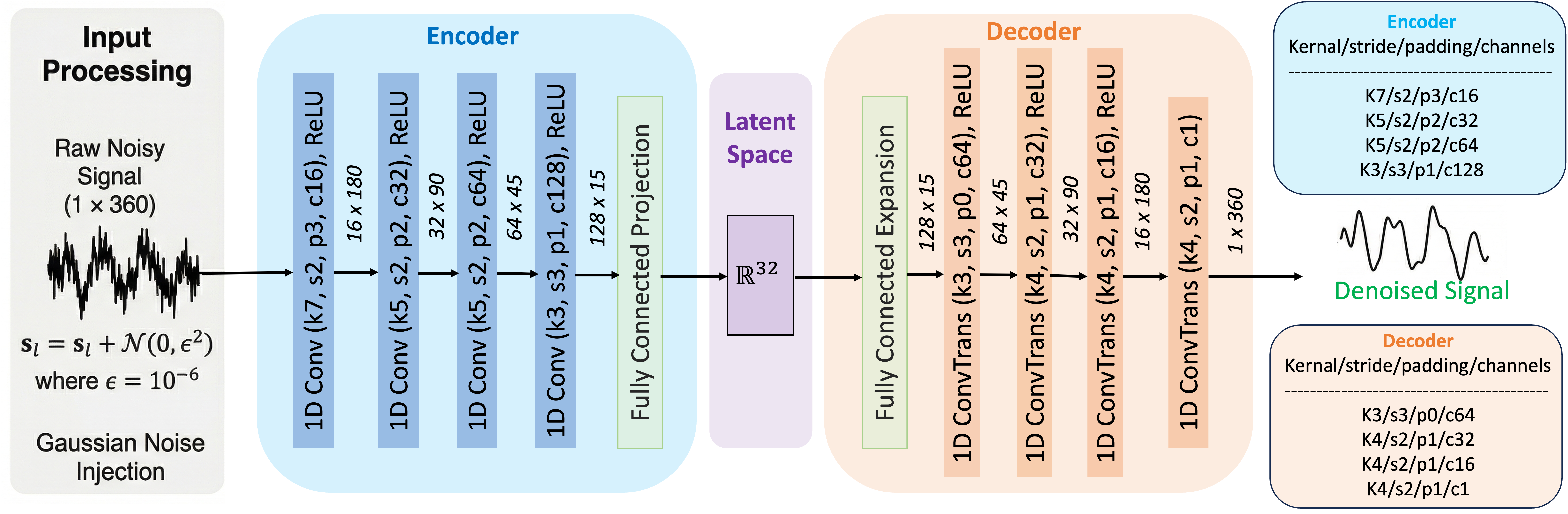}
    \caption{Architecture of the denoising network $\mathcal{N}_d$. The input Lagrangian signal is first perturbed with a small Gaussian noise term, $s_l = s_l + \mathcal{N}(0,\epsilon^2)$, to improve robustness during training. The encoder consists of four 1D convolutional layers followed by a fully connected projection to a 32-dimensional latent representation. The decoder reconstructs the denoised signal through a fully connected expansion and four transposed 1D convolutional layers arranged symmetrically to the encoder.}
    \label{fig:N_d}
\end{figure}

This section describes the architectural and training details of two neural networks used in \textit{ELITE-RR}: the skin-aware projection network ($\mathcal{N}_p$) and the denoising network ($\mathcal{N}_d$). To enable effective learning across representation despite temporal misalignment, we introduce a phase-independent contrastive spectral loss that emphasizes frequency consistency over phase coherence, facilitating cross-modal training.

\noindent
\textbf{$\mathbf{\mathcal{N}_p}$:}
The $\mathcal{N}_p$ takes as input a 12-dimensional skin tone-aware feature vector $\mathbf{f}_s$ (as described in Section \ref{skin-tone-vector}), and produces a projection vector $\mathbf{p}_j = [\alpha, \beta, \gamma]$ that linearly combines the RGB temporal signals into a single physiological rPPG signal, $\mathbf{s}_e$. The architecture of $\mathcal{N}_p$ is a shallow fully connected network composed of three linear layers with ReLU activation:
(1) Input layer: $\mathbb{R}^{12} \rightarrow \mathbb{R}^{16}$
(2) Hidden layer: $\mathbb{R}^{16} \rightarrow \mathbb{R}^{8}$
(3) Output layer: $\mathbb{R}^{8} \rightarrow \mathbb{R}^{3}$ (representing $[\alpha, \beta, \gamma]$).
No final activation function is used, allowing $\mathcal{N}_p$ to learn optimal projection weights without constraints. 

\noindent
\textbf{$\mathbf{\mathcal{N}_d}$:}
To improve the reliability of Lagrangian rPPG signals, we introduce a 1D convolutional autoencoder-based denoising network, $\mathcal{N}_d$. The $\mathcal{N}_d$ is designed to suppress non-respiratory artifacts and enhance the respiratory components of the signal. Before feeding the raw signal, $\mathbf{s}_l$ into $\mathcal{N}_d$, we inject a small amount of Gaussian noise: $\mathbf{s}_l = \mathbf{s}_l + \mathcal{N}(0, \epsilon^2)$, where $\epsilon$ is a small constant (set to $10^{-6}$ in our implementation). This noise injection serves as a regularization strategy to improve the network's robustness to signal perturbations and generalize better during training.

The architecture of $\mathcal{N}_d$ consists of three main components as shown in Figure \ref{fig:N_d}:
\textbf{1. Encoder:} composed of four 1D convolutional layers with progressively increasing channels (16 $\rightarrow$128), followed by a fully connected projection to the latent space ($\mathbb{R}^{128\times15} \rightarrow \mathbb{R}^{32}$).
\textbf{2. Decoder:} It reconstructs the denoised form of the input signal, $\mathbf{s}_l^d$, using a fully connected layer followed by four transposed convolutional layers, arranged to mirror encoder’s structure. 
\textbf{3. RR Estimator:} A separate fully connected regression head is attached to the latent representation to predict the RR, $rr_{lt}$, from the latent embedding. This auxiliary task encourages the network to preserve relevant respiratory features during denoising.

\noindent
\textbf{Training:}
To jointly train $\mathcal{N}_p$ and $\mathcal{N}_d$ networks, we define a composite loss function, $\mathcal{L}_t$, that promotes cross-modal consistency and physiological relevance. Since the Eulerian and Lagrangian representations capture respiratory information from different visual cues, our goal is to align them spectrally while preserving fidelity to the ground-truth respiration signal. To this end, $\mathcal{L}_t$ combines contrastive spectral losses and a data fidelity term, and is formulated as:
\begin{align*}
\mathcal{L}_t =\ & \lambda_1 \mathcal{L}_c(\mathbf{s}_e, \mathbf{s}_l^d) + \lambda_2 \mathcal{L}_c(\mathbf{s}_e, \mathbf{s}_{gt}) \\
& + \lambda_3 \mathcal{L}_c(\mathbf{s}_l^d, \mathbf{s}_{gt}) + \lambda_4 \mathcal{L}_{df}(rr_{lt}, rr_{gt})
\end{align*}
Here, $\mathcal{L}_c$ denotes the contrastive spectral loss, encouraging spectral alignment between (i) the Eulerian signal $\mathbf{s}_e$ and the denoised Lagrangian signal $\mathbf{s}_l^d$, and (ii) each of these signals with the ground truth signal $\mathbf{s}_{gt}$. This promotes representation-agnostic learning focused on consistent respiratory frequency content. The data fidelity term, $\mathcal{L}_{df}$, ensures that the auxiliary RR estimation from the latent space of $\mathcal{N}_d$ ($rr_{lt}$) closely matches the ground truth RR ($rr_{gt}$), encouraging frequency-aware latent encoding. The hyperparameters $\lambda_1$, $\lambda_2$, $\lambda_3$, and $\lambda_4$ control the relative importance of each loss and are selected based on validation performance.

\noindent
\textbf{Contrastive Spectral Loss:}
In dual-representation rPPG estimation, signals derived from Eulerian and Lagrangian representation often exhibit phase differences due to variations in their extraction mechanisms and noise characteristics, despite sharing similar frequency content. This phase misalignment makes traditional time-domain losses, such as Mean Squared Error (MSE), sub-optimal for supervision. In response, we propose a contrastive spectral loss, $\mathcal{L}_c$, that operates in the frequency domain and focuses on aligning respiratory frequency components, irrespective of their temporal phase. Let $x(t)$, $y(t)$ be two time-domain signals, representing the projected Eulerian and denoised Lagrangian rPPG signal. The loss is computed in the following steps: 

\noindent
\textbf{1. Frequency Transformation via FFT:}
We convert both signals into the frequency domain using the one-sided FFT, discarding the DC component to focus on oscillatory behavior: $\mathbf{X} = \text{FFT}(x)[1:], \quad \mathbf{Y} = \text{FFT}(y)[1:]$

\noindent
\textbf{2. Magnitude Spectrum Normalization:}
We compute the magnitude spectra and normalize them to form valid probability distributions:
$\mathbf{p} = \frac{|\mathbf{X}|}{(\sum |\mathbf{X}|) + \varepsilon}, \quad 
\mathbf{q} = \frac{|\mathbf{Y}|}{(\sum |\mathbf{Y}|) + \varepsilon}$,
where $\varepsilon$ is a small constant to avoid division by zero.

\noindent
\textbf{3. Cumulative Distribution Computation:}
We compute the cumulative distribution functions (CDFs) of the normalized spectra:
$\mathbf{P} = \text{cumsum}(\mathbf{p}), \quad 
\mathbf{Q} = \text{cumsum}(\mathbf{q})
$.

\noindent
\textbf{4. Earth Mover’s Distance Calculation:}
The $\mathcal{L}_c$ is defined as the $L_1$ distance between the two cumulative distributions, which corresponds to the Earth Mover’s Distance (EMD):
$
\mathcal{L}_c(x, y) = \left\| \mathbf{P} - \mathbf{Q} \right\|_1
$.
It quantifies the minimal cost to transform one spectral distribution into another, offering a phase-invariant and frequency-focused supervision signal. It is particularly suited for aligning signals that share similar dominant frequency components but may differ in phase.

\section{Experimental Analysis}
\label{sec:experimental}

\begin{figure}
    \centering
    \includegraphics[width=\linewidth]{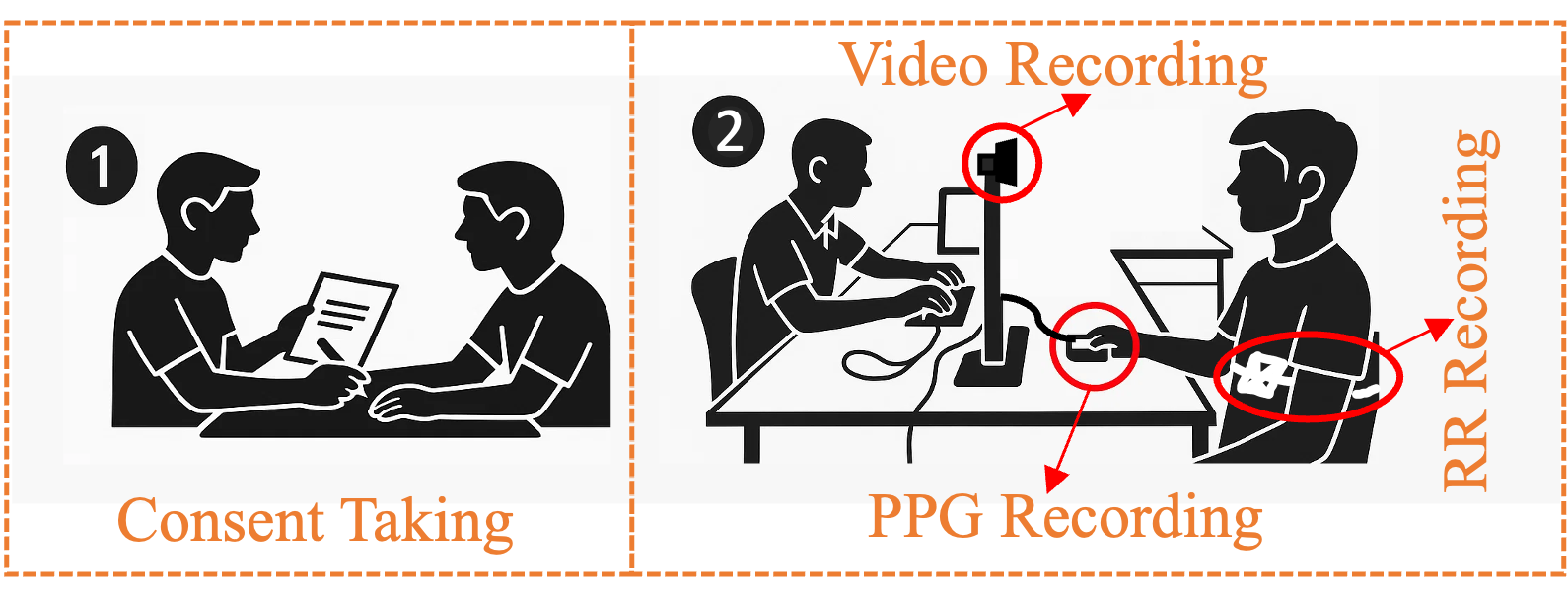}
    \caption{Overview of RR-rPPG dataset acquisition protocol.}
    \label{RR-rPPG_acquisition}
\end{figure}

\begin{figure}
    \centering
\includegraphics[width=\linewidth]{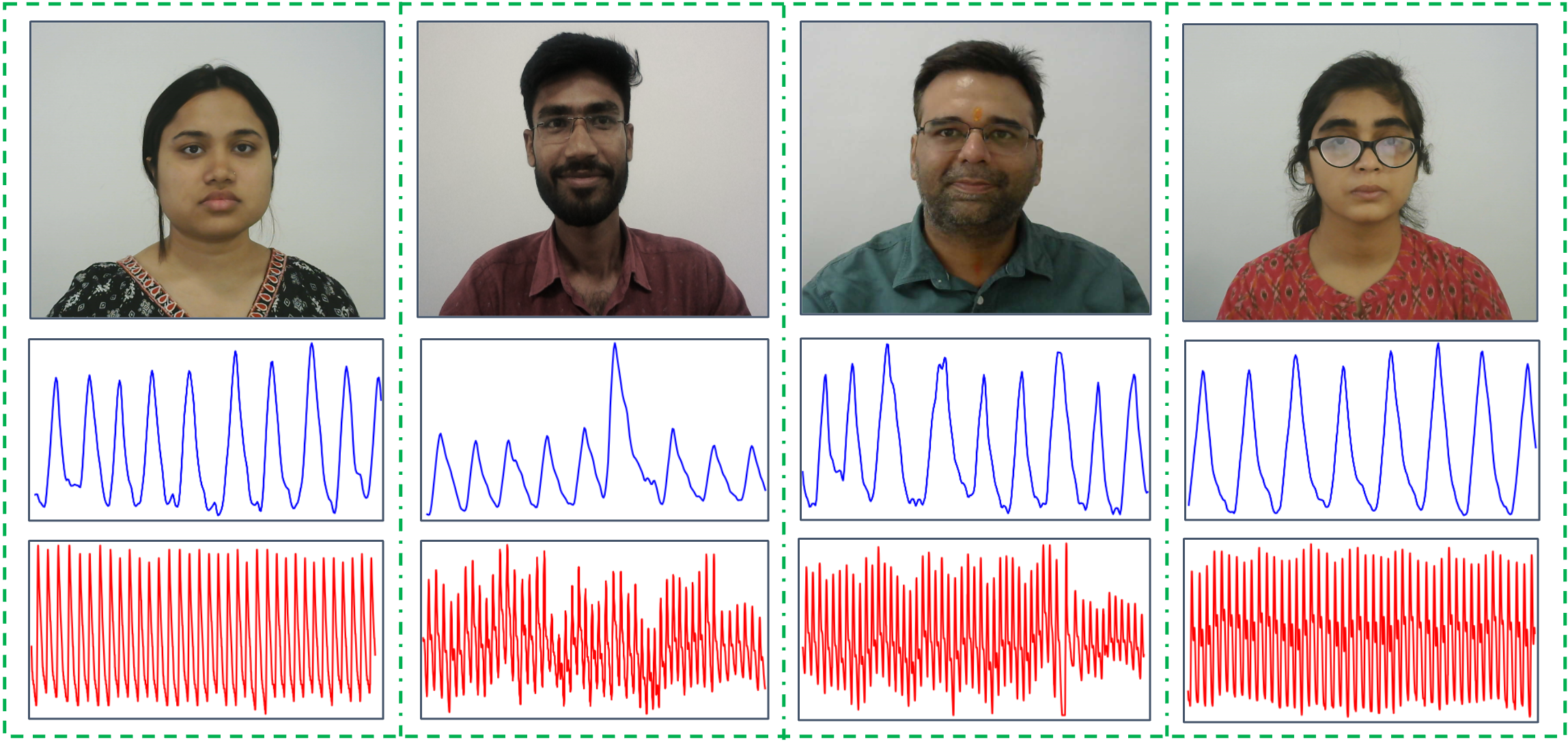}
    \caption{Representative samples from the proposed RR-rPPG dataset. Each column presents a subject’s facial frame (top) together with the corresponding ground-truth respiration signal (blue) and pulse signal (red) (bottom). The subject(s) in the figure gave permission and consent for the use of their identifiable images.}
    \label{fig:Samples_RR-rPPG}
\end{figure}

\subsection{Dataset Description}

\begin{figure}
    \centering
    \includegraphics[width=\linewidth]{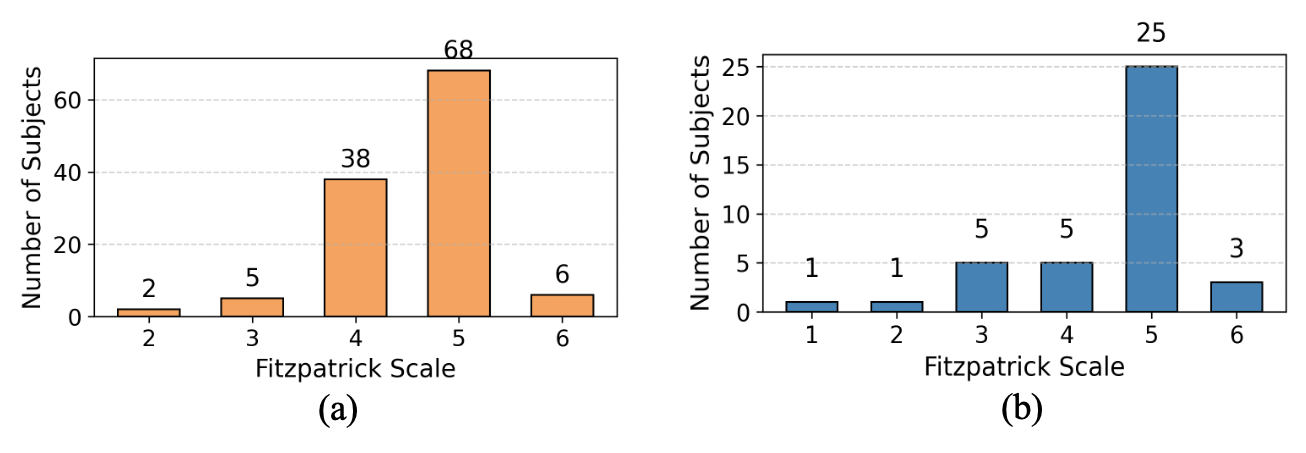}
    \caption{Fitzpatrick Skin-Tone Distribution in the (a) RR-rPPG and (b) COHFACE Datasets.}
    \label{fig:fitzpatrick}
\end{figure}

\noindent
\textit{\textbf{Proposed Dataset:}}
We introduce the RR-rPPG dataset, comprising 119 facial video recordings from participants aged between 18 and 40 years, including 26 females and 93 males. Alongside each video, corresponding ground truth respiratory and pulse signals were collected as shown in Figure \ref{RR-rPPG_acquisition}. Each recording session lasted for 90 seconds. 
Participants provided written informed consent after being briefed on the study protocol and risks. 
Demographic information, such as age and gender, was documented. Participants were seated approximately 1 meter from an RGB camera in a naturally lit environment. Facial videos were captured using an HP-W200 camera at a resolution of 640$\times$480 pixels and 30 frames per second (fps). Ground truth (GT) respiratory signals were obtained using a Go Direct Respiration Belt placed around the chest. This sensor measures respiratory effort in Newtons (N), with a 0-50 N measurement range and a resolution of 0.01 N. We used a Beurer PO80 pulse oximeter placed on the right index finger to record pulse signals. It provides pulse rate measurements within the range of 30-250 beats per minute (BPM), with an accuracy of $\pm$2 BPM. Additionally, each video and corresponding GT signals were timestamped to ensure precise temporal alignment and synchronization. Figure \ref{fig:Samples_RR-rPPG} shows sample data of our proposed dataset, RR-rPPG.

\noindent
\textbf{\textit{Existing Dataset:}}
We also use the COHFACE dataset \citep{heusch2017reproducible}, comprising 160, 1-minute facial videos (20 fps, 640×480) from 40 subjects, along with synchronized pulse and respiratory signals recorded using Thought Technology sensors. 

The RR-rPPG dataset has been specifically curated to capture a broad range of respiratory patterns across individuals with diverse Indian skin tones and facial features, ensuring greater representativeness and robustness for real-world rPPG-based applications. 
Table \ref{tab:my-table} summarizes the distribution of respiratory dynamics in the RR-rPPG and publicly available COHFACE datasets. The RR-rPPG dataset exhibits a wider range and higher variability in respiratory rates, which is valuable for training models to generalize across different breathing patterns. The skin-tone distributions of both datasets are illustrated in Figure \ref{fig:fitzpatrick}.

\begin{table}[!h]
\centering
\caption{Distribution of respiratory dynamics in the RR-rPPG and COHFACE datasets.}
\label{tab:my-table}
\begin{adjustbox}{width=0.65\linewidth}
\begin{threeparttable}
\begin{tabular}{|ccccc|}
\hline
\textbf{Datasets} & \textbf{Min}  & \textbf{Max}   & \textbf{Mean}  & \textbf{STD}  \\ \hline
RR-rPPG  & 5.00 & 26.00 & 17.00 & 5.80 \\ \hline
COHFACE  & 6.00    & 25.00    & 12.00    & 4.57 \\ \hline
\end{tabular}
\begin{tablenotes}
\item[Note] All parameters are reported in BrPM.
\end{tablenotes}
\end{threeparttable}
\end{adjustbox}
\end{table}

\subsection{Implementation Details} 
All ROI and signal extractions were done in MATLAB 2023a on an Intel® Core™ i7-9700 CPU. Training and evaluation were performed on a server with an Intel Xeon Gold 6240 CPU, 192GB RAM, and NVIDIA V100 GPU. The $\mathcal{N}_{p}$ and $\mathcal{N}_{d}$ networks were trained using the Adam optimizer for 30 epochs (batch size 16, learning rate 0.0001). 
Each video was segmented into non-overlapping 12-second clips, treated as individual samples. Details regarding this clip length choice are provided in the following subsection. For each dataset, 80\% of subjects were used for training, 10\% for validation, and 10\% for testing. Performance was evaluated using Mean Absolute Error (MAE), Root Mean Squared Error (RMSE), and Pearson correlation coefficient ($r$), reported in BrPM. These metrics were chosen to ensure consistency and comparability with prior works in the literature, where they are commonly used to assess RR estimation performance \citep{luguern2020assessment}. Lower MAE and RMSE, along with higher $r$, indicate better estimation accuracy.

\subsection{Hyperparameter Selection}
\begin{figure}
    \centering
    \includegraphics[width=0.8\linewidth]{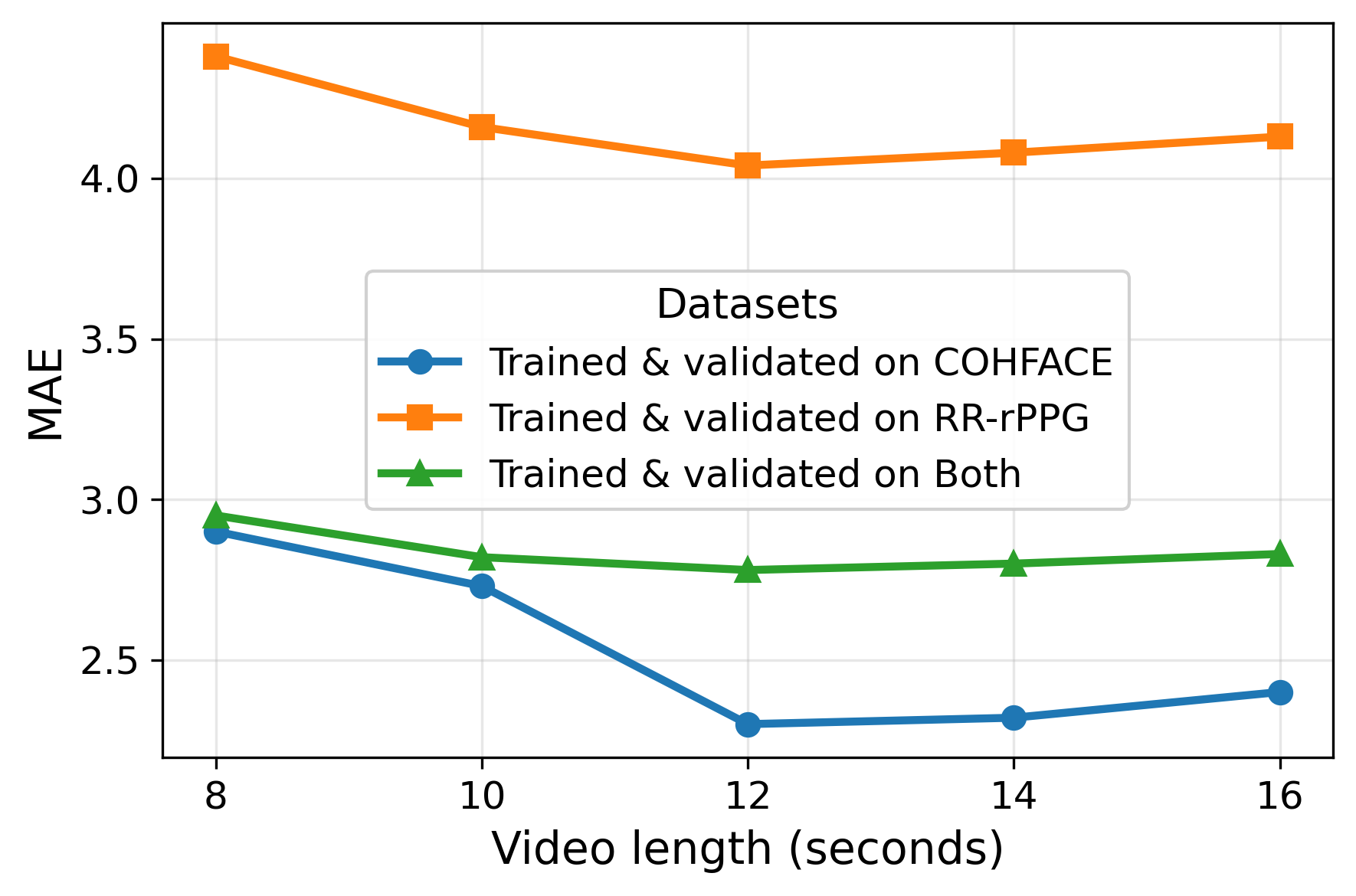}
    \caption{Performance of the proposed method across different video clip lengths, with optimal results observed at 12 seconds.}
    \label{fig:hyper}
\end{figure}

An important hyperparameter in \textit{ELITE-RR} is the length of the input video clip. To determine the optimal configuration, we evaluate the method’s performance across varying clip lengths (8 to 16 seconds), as shown in Figure~\ref{fig:hyper}. The results demonstrate that a 12-second clip length consistently yields better performance across all training settings.

We hypothesize that this is due to the physiological range of respiratory rate (RR) in healthy individuals, which typically spans from 5 to 30 BrPM \citep{vatanparvar2022respiration}.  A 12-second duration ensures that at least one full respiratory cycle is captured even for the lowest RR values. Additionally, compared to longer clips such as 14 or 16 seconds, the 12-second clips offer a balanced trade-off between capturing sufficient respiratory information and maintaining a higher number of training samples. As longer clips reduce the number of samples per video, they can negatively impact training. Consequently, this setting leads to improved generalization and more stable performance, and is adopted as the default configuration in all experiments.

\subsection{Comparative Analysis}
\begin{table}[]
\centering
\caption{Comparative analysis: \textit{ELITE-RR} with existing methods.}
\label{compare table}
\begin{adjustbox}{width=\columnwidth}
\begin{threeparttable}
\begin{tabular}{|c|ccc|ccc|ccc|}
\hline
\multirow{2}{*}{\textbf{Method}} & \multicolumn{3}{c|}{\textbf{COHFACE + RR-rPPG}} & \multicolumn{3}{c|}{\textbf{RR-rPPG}} & \multicolumn{3}{c|}{\textbf{COHFACE}} \\
\cline{2-10}
& \textbf{MAE} & \textbf{RMSE} & \textbf{r} & \textbf{MAE} & \textbf{RMSE} & \textbf{r} & \textbf{MAE} & \textbf{RMSE} & \textbf{r} \\
\hline
\citep{vatanparvar2022respiration}  & 7.74 & 9.62  & 0.08 & 11.79 & 12.69 & 0.31 & 6.71 & 7.53  & 0.15 \\ 
\citep{vatanparvar2022respiration}*           & 5.29 & 6.74  & 0.12  & 6.92  & 9.66  & 0.17 & 4.45 & 5.67  & 0.17 \\ \hline
\cite{seo2025estimation} & 7.98 & 9.59  & 0.12  & 10.94 & 12.16 & 0.05 & 5.24 & 6.18  & 0.13 \\
\citep{seo2025estimation}*     & 6.14 & 7.74  & 0.23  & 9.55  & 11.10  & 0.13 & 3.86 & 5.80  & 0.17 \\ \hline
\citep{luguern2020assessment}           & 6.90 & 8.84  & 0.11  & 8.76  & 9.13  & 0.16 & 4.41 & 5.29  & 0.13 \\ 
\citep{luguern2020assessment}*          & 5.10 & 6.33  & 0.14  & 6.89  & 7.92  & 0.19 & 3.99 & 4.56  & 0.15 \\ \hline
\citep{du2021weakly}       & 10.24 & 11.49 & 0.20  & 7.58  & 9.49  & 0.04 & 9.85 & 10.88 & 0.16 \\ 
\citep{du2021weakly}*       & 8.21 & 9.22  & 0.27  & 5.59  & 7.84  & 0.11 & 7.25 & 8.75  & 0.19 \\ \hline
\citep{gwak2024multimodal}     & 6.99 & 8.58  & 0.18  & 8.68  & 10.38 & 0.15 & 4.57 & 5.51  & 0.35 \\
\citep{gwak2024multimodal}*     & 4.37 & 7.00  & 0.21  & 6.98  & 8.88  & 0.17 & 3.71 & 4.65  & 0.36 \\ \hline
\textbf{\textit{ELITE-RR}}        & \textbf{2.79} & \textbf{3.34} & \textbf{0.25} & \textbf{4.04} & \textbf{4.74} & \textbf{0.44} & \textbf{2.31} & \textbf{3.12} & \textbf{0.39} \\
\hline
\end{tabular}
\begin{tablenotes}
\item[$*$] These methods use our proposed skin-tone-aware rPPG signal extraction in place of their original rPPG extraction technique.
\end{tablenotes}
\end{threeparttable}
\end{adjustbox}
\end{table}

We compare our proposed method, \textit{ELITE-RR}, against existing rPPG-based RR estimation methods \citep{vatanparvar2022respiration, luguern2020assessment, du2021weakly, gwak2024multimodal, seo2025estimation} on RR-rPPG and COHFACE datasets. Results are summarized in Table \ref{compare table}. All existing methods were implemented based on their published descriptions. Notably, we excluded the domain classifier from \citep{du2021weakly} due to the absence of domain labels and the unavailability of their original dataset. Additionally, we report the performance of these methods when their Eulerian signal extraction is replaced with our skin-tone-aware rPPG signal formulation.

The Table \ref{compare table} demonstrates that \textit{ELITE-RR} consistently outperforms existing methods across both individual datasets and the combined evaluation setting. Notably, incorporating our skin-tone-aware signal extraction into existing methods (\textit{denoted by *}) leads to marked improvements in their performance, underscoring the effectiveness of our projection strategy. \textit{ELITE-RR} achieves the lowest MAE and RMSE, along with the highest $r$, demonstrating superior performance for RR estimation. These highlight the value of our unified dual-representation design and spectral alignment strategy for reliable RR estimation. 

\subsection{Ablation Study}
\begin{table}[]
\centering
\caption{Analysis of the importance of different components of the proposed method, \textit{ELITE-RR} across different datasets.}
\label{tab:analysis-of-components}
\begin{adjustbox}{width=\linewidth}
\begin{threeparttable}
\begin{tabular}{|c|ccccccccc|}
\hline
\multirow{3}{*}{\textbf{Methods}}      & \multicolumn{9}{c|}{\textbf{Datasets}}                                                                                                                                                                                                         \\ \cline{2-10} 
                              & \multicolumn{3}{c|}{\textbf{COHFACE \& RR-rPPG}}                                            & \multicolumn{3}{c|}{\textbf{RR-rPPG}}                                                       & \multicolumn{3}{c|}{\textbf{COHFACE}}                                 \\ \cline{2-10} 
                              & \multicolumn{1}{c|}{\textbf{MAE}}  & \multicolumn{1}{c|}{\textbf{RMSE}} & \multicolumn{1}{c|}{\textbf{r}}    & \multicolumn{1}{c|}{MAE}  & \multicolumn{1}{c|}{\textbf{RMSE}} & \multicolumn{1}{c|}{\textbf{r}}     & \multicolumn{1}{c|}{\textbf{MAE}}  & \multicolumn{1}{c|}{\textbf{RMSE}} & \textbf{r}    \\ \hline
\textit{FusionRR-mse}          & \multicolumn{1}{c|}{4.17} & \multicolumn{1}{c|}{5.09} & \multicolumn{1}{c|}{0.15} & \multicolumn{1}{c|}{4.53} & \multicolumn{1}{c|}{5.36} & \multicolumn{1}{c|}{0.12}  & \multicolumn{1}{c|}{2.70} & \multicolumn{1}{c|}{3.44} & 0.34 \\ 
\textit{FusionRR-np}            & \multicolumn{1}{c|}{3.89} & \multicolumn{1}{c|}{4.80} & \multicolumn{1}{c|}{0.13} & \multicolumn{1}{c|}{4.75} & \multicolumn{1}{c|}{5.99} & \multicolumn{1}{c|}{-0.01} & \multicolumn{1}{c|}{2.61} & \multicolumn{1}{c|}{3.36} & 0.21 \\ \hline
\textit{FusionRR-wd}               & \multicolumn{1}{c|}{3.75} & \multicolumn{1}{c|}{4.71} & \multicolumn{1}{c|}{0.08} & \multicolumn{1}{c|}{4.49} & \multicolumn{1}{c|}{5.80} & \multicolumn{1}{c|}{-0.15} & \multicolumn{1}{c|}{2.77} & \multicolumn{1}{c|}{4.09} & 0.15 \\ 

\textit{Lagrangian}         & \multicolumn{1}{c|}{3.65} & \multicolumn{1}{c|}{4.01} & \multicolumn{1}{c|}{0.21} & \multicolumn{1}{c|}{4.41} & \multicolumn{1}{c|}{5.22} & \multicolumn{1}{c|}{0.17} & \multicolumn{1}{c|}{2.66} & \multicolumn{1}{c|}{3.14} & 0.26 \\ \hline

\textit{FusionRR-red}           & \multicolumn{1}{c|}{5.26} & \multicolumn{1}{c|}{6.11} & \multicolumn{1}{c|}{0.01} & \multicolumn{1}{c|}{6.22} & \multicolumn{1}{c|}{6.29} & \multicolumn{1}{c|}{0.15}  & \multicolumn{1}{c|}{3.16} & \multicolumn{1}{c|}{4.43} & -0.23 \\ 
\textit{FusionRR-green}           & \multicolumn{1}{c|}{4.93} & \multicolumn{1}{c|}{5.50} & \multicolumn{1}{c|}{-0.16} & \multicolumn{1}{c|}{5.70} & \multicolumn{1}{c|}{5.72} & \multicolumn{1}{c|}{-0.14}  & \multicolumn{1}{c|}{2.96} & \multicolumn{1}{c|}{3.87} & 0.16 \\ 
\textit{FusionRR-blue}          & \multicolumn{1}{c|}{4.87} & \multicolumn{1}{c|}{5.60} & \multicolumn{1}{c|}{-0.01} & \multicolumn{1}{c|}{5.44} & \multicolumn{1}{c|}{5.93} & \multicolumn{1}{c|}{-0.10}  & \multicolumn{1}{c|}{2.69} & \multicolumn{1}{c|}{3.78} & -0.24 \\ 
\textit{FusionRR-Chrom}          & \multicolumn{1}{c|}{5.26} & \multicolumn{1}{c|}{6.11} & \multicolumn{1}{c|}{-0.11} & \multicolumn{1}{c|}{5.72} & \multicolumn{1}{c|}{5.77} & \multicolumn{1}{c|}{0.11}  & \multicolumn{1}{c|}{3.17} & \multicolumn{1}{c|}{4.43} & 0.04 \\ \hline

\textit{Eulerian-avg}                & \multicolumn{1}{c|}{4.17} & \multicolumn{1}{c|}{5.16} & \multicolumn{1}{c|}{0.11} & \multicolumn{1}{c|}{4.73} & \multicolumn{1}{c|}{5.99} & \multicolumn{1}{c|}{0.13} & \multicolumn{1}{c|}{2.79} & \multicolumn{1}{c|}{3.47} & 0.25 \\ 
\textit{Eulerian-BSNR}             & \multicolumn{1}{c|}{4.13} & \multicolumn{1}{c|}{5.04} & \multicolumn{1}{c|}{0.14} & \multicolumn{1}{c|}{4.70} & \multicolumn{1}{c|}{4.95} & \multicolumn{1}{c|}{0.16}  & \multicolumn{1}{c|}{2.52} & \multicolumn{1}{c|}{3.58} & 0.23 \\ 
\textit{Eulerian-BSD}              & \multicolumn{1}{c|}{4.20} & \multicolumn{1}{c|}{5.08} & \multicolumn{1}{c|}{0.18} & \multicolumn{1}{c|}{4.92} & \multicolumn{1}{c|}{5.23} & \multicolumn{1}{c|}{0.14} & \multicolumn{1}{c|}{2.66} & \multicolumn{1}{c|}{3.74} & 0.17 \\ 
\textit{Eulerian-WSNR}        & \multicolumn{1}{c|}{4.15} & \multicolumn{1}{c|}{5.13} & \multicolumn{1}{c|}{0.15} & \multicolumn{1}{c|}{4.71} & \multicolumn{1}{c|}{4.97} & \multicolumn{1}{c|}{0.12} & \multicolumn{1}{c|}{2.59} & \multicolumn{1}{c|}{3.47} & 0.25 \\ 
\textit{Eulerian-WSD}         & \multicolumn{1}{c|}{4.18} & \multicolumn{1}{c|}{5.10} & \multicolumn{1}{c|}{0.16} & \multicolumn{1}{c|}{4.74} & \multicolumn{1}{c|}{4.98} & \multicolumn{1}{c|}{0.13} & \multicolumn{1}{c|}{2.58} & \multicolumn{1}{c|}{3.49} & 0.24 \\ \hline
\textit{Multiview-BSNR}               & \multicolumn{1}{c|}{4.13} & \multicolumn{1}{c|}{5.04} & \multicolumn{1}{c|}{0.14} & \multicolumn{1}{c|}{4.71} & \multicolumn{1}{c|}{4.98} & \multicolumn{1}{c|}{0.23}  & \multicolumn{1}{c|}{2.50} & \multicolumn{1}{c|}{3.46} & 0.33 \\ 
\textit{Multiview-BSD}                & \multicolumn{1}{c|}{4.08} & \multicolumn{1}{c|}{4.98} & \multicolumn{1}{c|}{0.15} & \multicolumn{1}{c|}{4.70} & \multicolumn{1}{c|}{4.95} & \multicolumn{1}{c|}{0.23}  & \multicolumn{1}{c|}{2.52} & \multicolumn{1}{c|}{3.48} & 0.33 \\ 
\textit{Multiview-WSD}            & \multicolumn{1}{c|}{3.62} & \multicolumn{1}{c|}{3.93} & \multicolumn{1}{c|}{0.18} & \multicolumn{1}{c|}{4.85} & \multicolumn{1}{c|}{5.12} & \multicolumn{1}{c|}{0.24}  & \multicolumn{1}{c|}{2.51} & \multicolumn{1}{c|}{3.46} & 0.32 \\ 
\textit{\textbf{ELITE-RR}} & \multicolumn{1}{c|}{\textbf{2.79}} & \multicolumn{1}{c|}{\textbf{3.34}} & \multicolumn{1}{c|}{\textbf{0.25}} & \multicolumn{1}{c|}{\textbf{4.04}} & \multicolumn{1}{c|}{\textbf{4.74}} & \multicolumn{1}{c|}{\textbf{0.44}}  & \multicolumn{1}{c|}{\textbf{2.31}} & \multicolumn{1}{c|}{\textbf{3.12}} & \textbf{0.39} \\ \hline
\end{tabular}
\end{threeparttable}
\end{adjustbox}
\end{table}

\begin{table}[]
\centering
\caption{Cross-dataset analysis of our proposed method.}
\label{tab:cross-datasets}
\begin{adjustbox}{width=0.85\columnwidth}
\begin{threeparttable}
\begin{tabular}{|c|c|ccc|}
\hline
\textbf{Train Dataset}                      & \textbf{Test Dataset} & \textbf{MAE}  & \textbf{RMSE} & {$\mathbf{r}$}   \\ \hline
COHFACE                            & RR-rPPG      & 4.89 & 6.05 & 0.02 \\ 
RR-rPPG                            & COHFACE  &   2.45 & 3.23 & 0.20 \\ \hline
\multirow{2}{*}{RR-rPPG \& COHFACE} & RR-rPPG      & \textbf{3.59} & \textbf{4.14} & \textbf{0.17} \\ \cline{2-5} 
                                   & COHFACE      & \textbf{2.39} & \textbf{3.19} & \textbf{0.38} \\ \hline
\end{tabular}
\end{threeparttable}
\end{adjustbox}
\end{table}
To assess the contribution of each component in our method, we conduct an ablation study summarized in Table \ref{tab:analysis-of-components}. We evaluate the impact of our $\mathcal{L}_c$ loss by replacing it with MSE (\textit{FusionRR-mse}) and negative Pearson loss (\textit{FusionRR-np}). We assess our $\mathcal{N}_d$ network’s role using \textit{FusionRR-wd}, which bypasses denoising. To analyze the effectiveness of our skin-aware rPPG extraction, we substitute it with red, green, blue, and CHROM signals in \textit{FusionRR-red}, \textit{FusionRR-green}, \textit{FusionRR-blue}, and \textit{FusionRR-Chrom}, respectively. For patch selection, we compare five Eulerian-only methods: \textit{Eulerian-avg} (uniform averaging), \textit{Eulerian-BSNR} (based on best SNR) and \textit{Eulerian-WSNR} (based on weighted SNR), and \textit{Eulerian-BSD} and \textit{Eulerian-WSD} (based on best and weighted quality metrics from \citep{gupta2023radiant}). We also include the Lagrangian-only method \textit{Lagrangian}. Finally, to analyze dual-representation fusion, we introduce \textit{Multiview-BSNR}, \textit{Multiview-BSD}, and \textit{Multiview-WSD}, which fuse Eulerian and Lagrangian predictions using SNR or the quality metric from \citep{gupta2023radiant}.

The Table \ref{tab:analysis-of-components} reveals that \textit{ELITE-RR} outperforms all variants across all datasets, achieving the lowest MAE and RMSE and the highest $r$. Replacing the contrastive loss with MSE or negative Pearson (\textit{FusionRR-mse}, \textit{FusionRR-np}) leads to noticeable performance drops, confirming the effectiveness of our spectral contrastive loss. Similarly, omitting the denoising network (\textit{FusionRR-wd}) results in degraded performance, highlighting the benefit of our denoising network. Replacing our Eulerian formulation with standard color-based signals (\textit{FusionRR-red/green/blue/Chrom}) significantly lowers performance, validating our skin-tone-aware strategy. Among patch selection techniques, best-patch-only selection based on SNR (\textit{Eulerian-BSNR}) outperforms simple averaging (\textit{Eulerian-avg}), \textit{Eulerian-BSD}, and also signal-weighted approaches (\textit{Eulerian-WSNR}, \textit{Eulerian-WSD}). Our proposed dual-representation fusion variant (\textit{ELITE-RR}) outperforms single-representation methods, confirming that combining Eulerian and Lagrangian cues enhances estimation performance. These results collectively demonstrate the effectiveness and necessity of each proposed component.

\begin{table*}[h]
\centering
\caption{Performance of signal processing–based respiratory rate estimation across Eulerian and Lagrangian modalities, using various blind source separation (BSS) techniques and different video clip lengths on the RR-rPPG dataset.}
\label{tab:sp-RR-RR}
\begin{adjustbox}{width=\textwidth}
\begin{threeparttable}
\begin{tabular}{|c|c|cccccccccccc|ccc|}
\hline
\multirow{3}{*}{\textbf{BSS Technique}} & \multirow{3}{*}{\begin{tabular}[c]{@{}c@{}}\textbf{Video Clip} \\ \textbf{Length}\\ \textbf{(seconds)}\end{tabular}} & \multicolumn{12}{c|}{\textbf{Eulerian Modality}}                                              & \multicolumn{3}{c|}{\textbf{Lagrangian Modality}}                                                                    \\ \cline{3-17} 
                               &                                                                                           & \multicolumn{3}{c|}{\textbf{Red}}                                                             & \multicolumn{3}{c|}{\textbf{Green}}                                                           & \multicolumn{3}{c|}{\textbf{Blue}}                                                            & \multicolumn{3}{c|}{\textbf{CHROM}}                                     & \multicolumn{1}{c|}{\multirow{2}{*}{\textbf{MAE}}} & \multicolumn{1}{c|}{\multirow{2}{*}{\textbf{RMSE}}} & \multirow{2}{*}{\textbf{r}} \\ \cline{3-14}
                               &                                                                                           & \multicolumn{1}{c|}{\textbf{MAE}}   & \multicolumn{1}{c|}{\textbf{RMSE}}  & \multicolumn{1}{c|}{\textbf{r}}     & \multicolumn{1}{c|}{\textbf{MAE}}   & \multicolumn{1}{c|}{\textbf{RMSE}}  & \multicolumn{1}{c|}{\textbf{r}}     & \multicolumn{1}{c|}{\textbf{MAE}}   & \multicolumn{1}{c|}{\textbf{RMSE}}  & \multicolumn{1}{c|}{\textbf{r}}     & \multicolumn{1}{c|}{\textbf{MAE}}  & \multicolumn{1}{c|}{\textbf{RMSE}}  & \textbf{r}     & \multicolumn{1}{c|}{}                     & \multicolumn{1}{c|}{}                      &                    \\ \hline
\multirow{6}{*}{PCA}           & 8                                                                                         & \multicolumn{1}{c|}{8.83}  & \multicolumn{1}{c|}{9.95}  & \multicolumn{1}{c|}{-0.01} & \multicolumn{1}{c|}{8.82}  & \multicolumn{1}{c|}{9.94}  & \multicolumn{1}{c|}{-0.01} & \multicolumn{1}{c|}{8.82}  & \multicolumn{1}{c|}{9.95}  & \multicolumn{1}{c|}{-0.01} & \multicolumn{1}{c|}{8.83} & \multicolumn{1}{c|}{9.95}  & 0.07  & \multicolumn{1}{c|}{5.53}                 & \multicolumn{1}{c|}{7.26}                  & 0.06               \\  
                               & 10                                                                                        & \multicolumn{1}{c|}{8.19}  & \multicolumn{1}{c|}{9.36}  & \multicolumn{1}{c|}{-0.03} & \multicolumn{1}{c|}{8.22}  & \multicolumn{1}{c|}{9.37}  & \multicolumn{1}{c|}{-0.04} & \multicolumn{1}{c|}{8.18}  & \multicolumn{1}{c|}{9.35}  & \multicolumn{1}{c|}{-0.01} & \multicolumn{1}{c|}{8.20} & \multicolumn{1}{c|}{9.36}  & 0.01  & \multicolumn{1}{c|}{5.47}                 & \multicolumn{1}{c|}{7.06}                  & 0.08               \\ 
                               & 12                                                                                        & \multicolumn{1}{c|}{7.04}  & \multicolumn{1}{c|}{8.23}  & \multicolumn{1}{c|}{0.02}  & \multicolumn{1}{c|}{7.18}  & \multicolumn{1}{c|}{8.38}  & \multicolumn{1}{c|}{-0.03} & \multicolumn{1}{c|}{7.03}  & \multicolumn{1}{c|}{8.22}  & \multicolumn{1}{c|}{-0.06} & \multicolumn{1}{c|}{6.86} & \multicolumn{1}{c|}{8.01}  & 0.02  & \multicolumn{1}{c|}{5.01}                 & \multicolumn{1}{c|}{6.48}                  & 0.12               \\ 
                               & 14                                                                                        & \multicolumn{1}{c|}{9.41}  & \multicolumn{1}{c|}{10.56} & \multicolumn{1}{c|}{-0.09} & \multicolumn{1}{c|}{9.39}  & \multicolumn{1}{c|}{10.54} & \multicolumn{1}{c|}{-0.03} & \multicolumn{1}{c|}{9.37}  & \multicolumn{1}{c|}{10.52} & \multicolumn{1}{c|}{0.01}  & \multicolumn{1}{c|}{9.37} & \multicolumn{1}{c|}{10.52} & 0.01  & \multicolumn{1}{c|}{5.59}                 & \multicolumn{1}{c|}{7.42}                  & 0.08               \\  
                               & 16                                                                                        & \multicolumn{1}{c|}{8.92}  & \multicolumn{1}{c|}{10.05} & \multicolumn{1}{c|}{0.01}  & \multicolumn{1}{c|}{8.93}  & \multicolumn{1}{c|}{10.05} & \multicolumn{1}{c|}{-0.01} & \multicolumn{1}{c|}{8.93}  & \multicolumn{1}{c|}{10.04} & \multicolumn{1}{c|}{0.01}  & \multicolumn{1}{c|}{8.90} & \multicolumn{1}{c|}{10.02} & 0.04  & \multicolumn{1}{c|}{5.39}                 & \multicolumn{1}{c|}{7.72}                  & 0.11               \\  
                               & 60                                                                                        & \multicolumn{1}{c|}{9.70}  & \multicolumn{1}{c|}{10.77} & \multicolumn{1}{c|}{0.05}  & \multicolumn{1}{c|}{9.69}  & \multicolumn{1}{c|}{10.76} & \multicolumn{1}{c|}{-0.06} & \multicolumn{1}{c|}{9.69}  & \multicolumn{1}{c|}{10.77} & \multicolumn{1}{c|}{-0.09} & \multicolumn{1}{c|}{9.67} & \multicolumn{1}{c|}{10.76} & -0.09 & \multicolumn{1}{c|}{5.76}                 & \multicolumn{1}{c|}{7.73}                  & 0.16               \\ \hline
\multirow{6}{*}{ICA}           & 8                                                                                         & \multicolumn{1}{c|}{10.30} & \multicolumn{1}{c|}{10.74} & \multicolumn{1}{c|}{-0.02} & \multicolumn{1}{c|}{10.52} & \multicolumn{1}{c|}{10.94} & \multicolumn{1}{c|}{0.01}  & \multicolumn{1}{c|}{10.43} & \multicolumn{1}{c|}{10.87} & \multicolumn{1}{c|}{-0.01} & \multicolumn{1}{c|}{9.69} & \multicolumn{1}{c|}{10.10} & 0.02  & \multicolumn{1}{c|}{8.05}                 & \multicolumn{1}{c|}{9.17}                  & 0.01               \\  
                               & 10                                                                                        & \multicolumn{1}{c|}{9.95}  & \multicolumn{1}{c|}{10.68} & \multicolumn{1}{c|}{0.01}  & \multicolumn{1}{c|}{9.92}  & \multicolumn{1}{c|}{10.94} & \multicolumn{1}{c|}{0.01}  & \multicolumn{1}{c|}{10.40} & \multicolumn{1}{c|}{10.42} & \multicolumn{1}{c|}{-0.08} & \multicolumn{1}{c|}{9.38} & \multicolumn{1}{c|}{10.34} & -0.01 & \multicolumn{1}{c|}{8.64}                 & \multicolumn{1}{c|}{9.42}                  & -0.01              \\ 
                               & 12                                                                                        & \multicolumn{1}{c|}{9.49}  & \multicolumn{1}{c|}{10.69} & \multicolumn{1}{c|}{-0.04} & \multicolumn{1}{c|}{9.11}  & \multicolumn{1}{c|}{10.11} & \multicolumn{1}{c|}{-0.06} & \multicolumn{1}{c|}{9.69}  & \multicolumn{1}{c|}{9.79}  & \multicolumn{1}{c|}{0.01}  & \multicolumn{1}{c|}{9.19} & \multicolumn{1}{c|}{10.16} & -0.06 & \multicolumn{1}{c|}{7.61}                 & \multicolumn{1}{c|}{7.96}                  & 0.07               \\ 
                               & 14                                                                                        & \multicolumn{1}{c|}{9.28}  & \multicolumn{1}{c|}{10.56} & \multicolumn{1}{c|}{0.06}  & \multicolumn{1}{c|}{9.26}  & \multicolumn{1}{c|}{10.25} & \multicolumn{1}{c|}{0.02}  & \multicolumn{1}{c|}{9.47}  & \multicolumn{1}{c|}{10.53} & \multicolumn{1}{c|}{0.01}  & \multicolumn{1}{c|}{9.06} & \multicolumn{1}{c|}{10.51} & 0.11  & \multicolumn{1}{c|}{7.63}                 & \multicolumn{1}{c|}{8.13}                  & 0.02               \\ 
                               & 16                                                                                        & \multicolumn{1}{c|}{8.76}  & \multicolumn{1}{c|}{10.48} & \multicolumn{1}{c|}{0.12}  & \multicolumn{1}{c|}{9.60}  & \multicolumn{1}{c|}{10.97} & \multicolumn{1}{c|}{0.06}  & \multicolumn{1}{c|}{9.93}  & \multicolumn{1}{c|}{10.87} & \multicolumn{1}{c|}{0.03}  & \multicolumn{1}{c|}{9.41} & \multicolumn{1}{c|}{10.72} & -0.06 & \multicolumn{1}{c|}{7.49}                 & \multicolumn{1}{c|}{8.08}                  & 0.08               \\  
                               & 60                                                                                        & \multicolumn{1}{c|}{8.56}  & \multicolumn{1}{c|}{10.07} & \multicolumn{1}{c|}{0.16}  & \multicolumn{1}{c|}{8.92}  & \multicolumn{1}{c|}{9.96}  & \multicolumn{1}{c|}{0.16}  & \multicolumn{1}{c|}{8.61}  & \multicolumn{1}{c|}{10.01} & \multicolumn{1}{c|}{0.10}  & \multicolumn{1}{c|}{8.61} & \multicolumn{1}{c|}{10.26} & 0.05  & \multicolumn{1}{c|}{7.33}                 & \multicolumn{1}{c|}{8.03}                  & 0.02               \\ \hline
\multirow{6}{*}{MUK}           & 8                                                                                         & \multicolumn{1}{c|}{8.65}  & \multicolumn{1}{c|}{9.83}  & \multicolumn{1}{c|}{-0.02} & \multicolumn{1}{c|}{8.64}  & \multicolumn{1}{c|}{9.82}  & \multicolumn{1}{c|}{-0.03} & \multicolumn{1}{c|}{8.68}  & \multicolumn{1}{c|}{9.90}  & \multicolumn{1}{c|}{-0.07} & \multicolumn{1}{c|}{7.35} & \multicolumn{1}{c|}{8.66}  & -0.01 & \multicolumn{1}{c|}{5.53}                 & \multicolumn{1}{c|}{7.23}                  & 0.09               \\ 
                               & 10                                                                                        & \multicolumn{1}{c|}{8.01}  & \multicolumn{1}{c|}{9.25}  & \multicolumn{1}{c|}{0.01}  & \multicolumn{1}{c|}{8.16}  & \multicolumn{1}{c|}{9.38}  & \multicolumn{1}{c|}{-0.05} & \multicolumn{1}{c|}{8.00}  & \multicolumn{1}{c|}{9.20}  & \multicolumn{1}{c|}{0.01}  & \multicolumn{1}{c|}{7.98} & \multicolumn{1}{c|}{9.16}  & 0.04  & \multicolumn{1}{c|}{5.52}                 & \multicolumn{1}{c|}{7.11}                  & 0.05               \\  
                               & 12                                                                                        & \multicolumn{1}{c|}{7.59}  & \multicolumn{1}{c|}{8.96}  & \multicolumn{1}{c|}{-0.05} & \multicolumn{1}{c|}{7.39}  & \multicolumn{1}{c|}{8.73}  & \multicolumn{1}{c|}{-0.01} & \multicolumn{1}{c|}{7.39}  & \multicolumn{1}{c|}{8.65}  & \multicolumn{1}{c|}{0.02}  & \multicolumn{1}{c|}{7.35} & \multicolumn{1}{c|}{8.66}  & -0.01 & \multicolumn{1}{c|}{5.53}                 & \multicolumn{1}{c|}{6.68}                  & 0.14               \\  
                               & 14                                                                                        & \multicolumn{1}{c|}{9.06}  & \multicolumn{1}{c|}{10.30} & \multicolumn{1}{c|}{.06}   & \multicolumn{1}{c|}{9.32}  & \multicolumn{1}{c|}{10.53} & \multicolumn{1}{c|}{-0.07} & \multicolumn{1}{c|}{9.12}  & \multicolumn{1}{c|}{10.44} & \multicolumn{1}{c|}{-0.04} & \multicolumn{1}{c|}{9.23} & \multicolumn{1}{c|}{10.46} & 0.06  & \multicolumn{1}{c|}{6.00}                 & \multicolumn{1}{c|}{7.64}                  & 0.06               \\  
                               & 16                                                                                        & \multicolumn{1}{c|}{9.15}  & \multicolumn{1}{c|}{10.39} & \multicolumn{1}{c|}{-0.07} & \multicolumn{1}{c|}{8.92}  & \multicolumn{1}{c|}{10.20} & \multicolumn{1}{c|}{0.04}  & \multicolumn{1}{c|}{9.03}  & \multicolumn{1}{c|}{10.28} & \multicolumn{1}{c|}{0.01}  & \multicolumn{1}{c|}{9.15} & \multicolumn{1}{c|}{10.34} & -0.02 & \multicolumn{1}{c|}{5.85}                 & \multicolumn{1}{c|}{7.60}                  & 0.09               \\ 
                               & 60                                                                                        & \multicolumn{1}{c|}{9.78}  & \multicolumn{1}{c|}{10.95} & \multicolumn{1}{c|}{0.08}  & \multicolumn{1}{c|}{9.52}  & \multicolumn{1}{c|}{10.60} & \multicolumn{1}{c|}{0.07}  & \multicolumn{1}{c|}{9.79}  & \multicolumn{1}{c|}{10.87} & \multicolumn{1}{c|}{-0.05} & \multicolumn{1}{c|}{9.73} & \multicolumn{1}{c|}{10.92} & -0.13 & \multicolumn{1}{c|}{5.74}                 & \multicolumn{1}{c|}{7.70}                  & 0.17               \\ \hline
\end{tabular}
\end{threeparttable}
\end{adjustbox}
\end{table*}

Furthermore, we perform a cross-dataset evaluation to assess the generalizability of \textit{ELITE-RR} between RR-rPPG and COHFACE (refer to Table \ref{tab:cross-datasets}). \textit{ELITE-RR} demonstrates improved generalizability when trained on both datasets, achieving the lowest MAE of 2.39 and the highest correlation of 0.38 on COHFACE. 

We also evaluate traditional signal processing–based RR estimation methods using three blind source separation (BSS) techniques: PCA, ICA, and Multiple Unit Kernel (MUK), across different video clip lengths on both RR-rPPG and COHFACE datasets. The results are presented in Table \ref{tab:sp-RR-RR} and Table \ref{tab:sp-RR-cohface}, respectively. These evaluations offer useful baselines for comparison with learning-based methods. Among all configurations, PCA with a 12-second clip length achieves relatively better performance across both representations and datasets.

\begin{figure}
    \centering
    \includegraphics[width=0.9\linewidth]{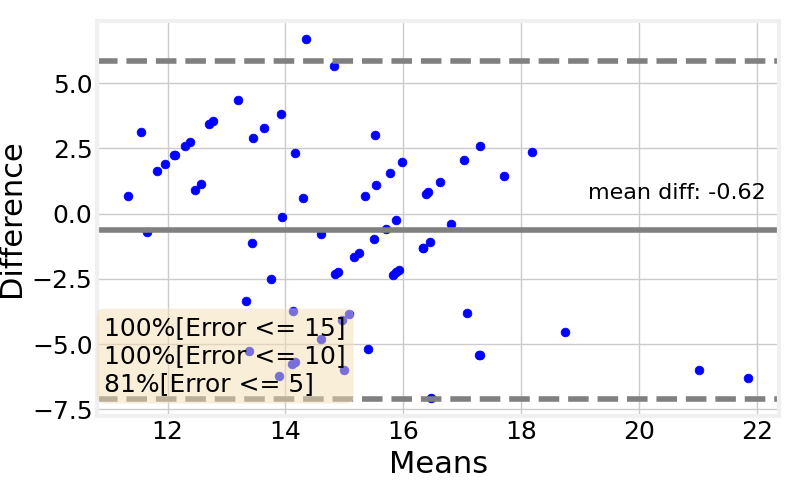}
    \caption{Bland--Altman plot of the proposed \textit{ELITE-RR} method on COHFACE + RR-rPPG dataset. The solid line represents the mean difference, while the dashed lines denote the limits of agreement.}
    \label{fig:ba_eliterr_combined}
\end{figure}

\subsection{Bland--Altman Analysis}
To further assess the agreement between the estimated RR and the ground-truth RR, we performed a Bland--Altman analysis for the proposed \textit{ELITE-RR} method on the combined COHFACE + RR-rPPG dataset. Figure \ref{fig:ba_eliterr_combined} shows the agreement between the estimated and ground-truth RR values by plotting their difference against their average for each sample.
The plot shows a mean difference of approximately $-0.62$ BrPM, indicating that, on average, ELITE-RR slightly underestimates the RR. Most samples lie within the limits of agreement, suggesting a reasonable level of consistency between the predicted and ground-truth values across a broad range of RR. In addition, a large proportion of the samples exhibit relatively small estimation errors, with approximately 81\% having an absolute error less than or equal to 5 BrPM, and all samples falling within 15 BrPM. These observations support the reliability of the proposed method for RR estimation.

\begin{table*}[]
\centering
\caption{Performance of signal processing–based RR estimation across Eulerian and Lagrangian representations, using various BSS techniques and different video clip lengths on the COHFACE dataset.}
\label{tab:sp-RR-cohface}
\begin{adjustbox}{width=\textwidth}
\begin{threeparttable}
\begin{tabular}{|c|c|cccccccccccc|ccc|}
\hline
\multirow{3}{*}{\textbf{BSS Technique}} & \multirow{3}{*}{\begin{tabular}[c]{@{}c@{}}\textbf{Video Clip} \\ \textbf{Length}\\ (\textbf{seconds})\end{tabular}} & \multicolumn{12}{c|}{\textbf{Eulerian Modality}}                                                                                                                                                                                                                                                                                      & \multicolumn{3}{c|}{\textbf{Lagrangian Modality}}                                                                    \\ \cline{3-17} 
                               &                                                                                           & \multicolumn{3}{c|}{\textbf{Red}}                                                           & \multicolumn{3}{c|}{\textbf{Green}}                                                         & \multicolumn{3}{c|}{\textbf{Blue}}                                                          & \multicolumn{3}{c|}{\textbf{CHROM}}                                    & \multicolumn{1}{c|}{\multirow{2}{*}{\textbf{MAE}}} & \multicolumn{1}{c|}{\multirow{2}{*}{\textbf{RMSE}}} & \multirow{2}{*}{\textbf{r}} \\ \cline{3-14}
                               &                                                                                           & \multicolumn{1}{c|}{\textbf{MAE}}  & \multicolumn{1}{c|}{\textbf{RMSE}} & \multicolumn{1}{c|}{\textbf{r}}     & \multicolumn{1}{c|}{\textbf{MAE}}  & \multicolumn{1}{c|}{\textbf{RMSE}} & \multicolumn{1}{c|}{\textbf{r}}     & \multicolumn{1}{c|}{\textbf{MAE}}  & \multicolumn{1}{c|}{\textbf{RMSE}} & \multicolumn{1}{c|}{\textbf{r}}     & \multicolumn{1}{c|}{\textbf{MAE}}  & \multicolumn{1}{c|}{\textbf{RMSE}} & \textbf{r}     & \multicolumn{1}{c|}{}                     & \multicolumn{1}{c|}{}                      &                    \\ \hline
\multirow{6}{*}{PCA}           & 8                                                                                         & \multicolumn{1}{c|}{4.22} & \multicolumn{1}{c|}{5.05} & \multicolumn{1}{c|}{-0.04} & \multicolumn{1}{c|}{4.20} & \multicolumn{1}{c|}{5.03} & \multicolumn{1}{c|}{-0.04} & \multicolumn{1}{c|}{4.11} & \multicolumn{1}{c|}{4.94} & \multicolumn{1}{c|}{0.01}  & \multicolumn{1}{c|}{4.07} & \multicolumn{1}{c|}{4.92} & 0.02  & \multicolumn{1}{c|}{4.86}                 & \multicolumn{1}{c|}{5.71}                  & 0.04               \\ 
                               & 10                                                                                        & \multicolumn{1}{c|}{4.11} & \multicolumn{1}{c|}{5.17} & \multicolumn{1}{c|}{-0.02} & \multicolumn{1}{c|}{4.03} & \multicolumn{1}{c|}{5.11} & \multicolumn{1}{c|}{-0.01} & \multicolumn{1}{c|}{4.13} & \multicolumn{1}{c|}{5.17} & \multicolumn{1}{c|}{-0.04} & \multicolumn{1}{c|}{3.89} & \multicolumn{1}{c|}{4.89} & 0.03  & \multicolumn{1}{c|}{4.59}                 & \multicolumn{1}{c|}{5.54}                  & 0.13               \\  
                               & 12                                                                                        & \multicolumn{1}{c|}{3.87} & \multicolumn{1}{c|}{4.89} & \multicolumn{1}{c|}{0.02}  & \multicolumn{1}{c|}{3.73} & \multicolumn{1}{c|}{4.69} & \multicolumn{1}{c|}{0.05}  & \multicolumn{1}{c|}{3.68} & \multicolumn{1}{c|}{4.64} & \multicolumn{1}{c|}{0.04}  & \multicolumn{1}{c|}{3.67} & \multicolumn{1}{c|}{4.61} & 0.04  & \multicolumn{1}{c|}{4.23}                 & \multicolumn{1}{c|}{5.33}                  & 0.08               \\  
                               & 14                                                                                        & \multicolumn{1}{c|}{4.04} & \multicolumn{1}{c|}{4.93} & \multicolumn{1}{c|}{0.03}  & \multicolumn{1}{c|}{3.95} & \multicolumn{1}{c|}{4.83} & \multicolumn{1}{c|}{0.07}  & \multicolumn{1}{c|}{3.99} & \multicolumn{1}{c|}{4.86} & \multicolumn{1}{c|}{0.04}  & \multicolumn{1}{c|}{3.97} & \multicolumn{1}{c|}{4.86} & -0.03 & \multicolumn{1}{c|}{4.87}                 & \multicolumn{1}{c|}{5.83}                  & 0.11               \\ 
                               & 16                                                                                        & \multicolumn{1}{c|}{4.36} & \multicolumn{1}{c|}{5.42} & \multicolumn{1}{c|}{0.01}  & \multicolumn{1}{c|}{4.32} & \multicolumn{1}{c|}{5.37} & \multicolumn{1}{c|}{0.05}  & \multicolumn{1}{c|}{4.32} & \multicolumn{1}{c|}{5.37} & \multicolumn{1}{c|}{-0.01} & \multicolumn{1}{c|}{4.44} & \multicolumn{1}{c|}{5.38} & -0.01 & \multicolumn{1}{c|}{5.16}                 & \multicolumn{1}{c|}{6.10}                  & 0.08               \\  
                               & 60                                                                                        & \multicolumn{1}{c|}{5.13} & \multicolumn{1}{c|}{6.34} & \multicolumn{1}{c|}{-0.19} & \multicolumn{1}{c|}{5.04} & \multicolumn{1}{c|}{6.13} & \multicolumn{1}{c|}{-0.01} & \multicolumn{1}{c|}{5.05} & \multicolumn{1}{c|}{6.13} & \multicolumn{1}{c|}{0.04}  & \multicolumn{1}{c|}{5.04} & \multicolumn{1}{c|}{6.13} & -0.02 & \multicolumn{1}{c|}{5.20}                 & \multicolumn{1}{c|}{6.44}                  & 0.20               \\ \hline
\multirow{6}{*}{ICA}           & 8                                                                                         & \multicolumn{1}{c|}{8.00} & \multicolumn{1}{c|}{8.96} & \multicolumn{1}{c|}{0.02}  & \multicolumn{1}{c|}{7.54} & \multicolumn{1}{c|}{7.75} & \multicolumn{1}{c|}{-0.01} & \multicolumn{1}{c|}{7.77} & \multicolumn{1}{c|}{8.64} & \multicolumn{1}{c|}{-0.03} & \multicolumn{1}{c|}{7.66} & \multicolumn{1}{c|}{8.55} & -0.01 & \multicolumn{1}{c|}{7.75}                 & \multicolumn{1}{c|}{8.03}                  & 0.03               \\ 
                               & 10                                                                                        & \multicolumn{1}{c|}{7.98} & \multicolumn{1}{c|}{8.81} & \multicolumn{1}{c|}{0.02}  & \multicolumn{1}{c|}{7.46} & \multicolumn{1}{c|}{7.72} & \multicolumn{1}{c|}{0.01}  & \multicolumn{1}{c|}{7.87} & \multicolumn{1}{c|}{8.31} & \multicolumn{1}{c|}{-0.04} & \multicolumn{1}{c|}{7.73} & \multicolumn{1}{c|}{8.66} & -0.04 & \multicolumn{1}{c|}{7.46}                 & \multicolumn{1}{c|}{8.14}                  & -0.04              \\ 
                               & 12                                                                                        & \multicolumn{1}{c|}{7.52} & \multicolumn{1}{c|}{8.46} & \multicolumn{1}{c|}{-0.07} & \multicolumn{1}{c|}{7.40} & \multicolumn{1}{c|}{7.47} & \multicolumn{1}{c|}{0.06}  & \multicolumn{1}{c|}{7.76} & \multicolumn{1}{c|}{8.15} & \multicolumn{1}{c|}{0.01}  & \multicolumn{1}{c|}{7.20} & \multicolumn{1}{c|}{8.29} & -0.08 & \multicolumn{1}{c|}{7.36}                 & \multicolumn{1}{c|}{7.62}                  & 0.04               \\  
                               & 14                                                                                        & \multicolumn{1}{c|}{7.61} & \multicolumn{1}{c|}{8.68} & \multicolumn{1}{c|}{-0.05} & \multicolumn{1}{c|}{8.13} & \multicolumn{1}{c|}{8.20} & \multicolumn{1}{c|}{-0.08} & \multicolumn{1}{c|}{7.89} & \multicolumn{1}{c|}{8.41} & \multicolumn{1}{c|}{-0.04} & \multicolumn{1}{c|}{7.72} & \multicolumn{1}{c|}{8.64} & -0.04 & \multicolumn{1}{c|}{8.65}                 & \multicolumn{1}{c|}{8.83}                  & 0.03               \\  
                               & 16                                                                                        & \multicolumn{1}{c|}{8.46} & \multicolumn{1}{c|}{8.76} & \multicolumn{1}{c|}{-0.01} & \multicolumn{1}{c|}{8.30} & \multicolumn{1}{c|}{8.84} & \multicolumn{1}{c|}{0.05}  & \multicolumn{1}{c|}{7.90} & \multicolumn{1}{c|}{8.34} & \multicolumn{1}{c|}{0.01}  & \multicolumn{1}{c|}{7.76} & \multicolumn{1}{c|}{8.89} & 0.01  & \multicolumn{1}{c|}{8.84}                 & \multicolumn{1}{c|}{8.91}                  & 0.14               \\ 
                               & 60                                                                                        & \multicolumn{1}{c|}{8.73} & \multicolumn{1}{c|}{9.46} & \multicolumn{1}{c|}{0.09}  & \multicolumn{1}{c|}{8.81} & \multicolumn{1}{c|}{9.74} & \multicolumn{1}{c|}{0.06}  & \multicolumn{1}{c|}{7.83} & \multicolumn{1}{c|}{8.58} & \multicolumn{1}{c|}{0.07}  & \multicolumn{1}{c|}{8.00} & \multicolumn{1}{c|}{9.93} & -0.05 & \multicolumn{1}{c|}{8.06}                 & \multicolumn{1}{c|}{8.51}                  & -0.03              \\ \hline
\multirow{6}{*}{MUK}           & 8                                                                                         & \multicolumn{1}{c|}{4.29} & \multicolumn{1}{c|}{5.47} & \multicolumn{1}{c|}{0.05}  & \multicolumn{1}{c|}{4.59} & \multicolumn{1}{c|}{5.65} & \multicolumn{1}{c|}{-0.06} & \multicolumn{1}{c|}{4.37} & \multicolumn{1}{c|}{5.70} & \multicolumn{1}{c|}{0.02}  & \multicolumn{1}{c|}{4.37} & \multicolumn{1}{c|}{5.38} & -0.01 & \multicolumn{1}{c|}{4.99}                 & \multicolumn{1}{c|}{5.85}                  & 0.04               \\ 
                               & 10                                                                                        & \multicolumn{1}{c|}{4.81} & \multicolumn{1}{c|}{6.04} & \multicolumn{1}{c|}{0.02}  & \multicolumn{1}{c|}{4.66} & \multicolumn{1}{c|}{5.81} & \multicolumn{1}{c|}{0.01}  & \multicolumn{1}{c|}{4.50} & \multicolumn{1}{c|}{5.61} & \multicolumn{1}{c|}{-0.02} & \multicolumn{1}{c|}{4.51} & \multicolumn{1}{c|}{5.73} & 0.01  & \multicolumn{1}{c|}{4.75}                 & \multicolumn{1}{c|}{5.69}                  & 0.14               \\ 
                               & 12                                                                                        & \multicolumn{1}{c|}{4.27} & \multicolumn{1}{c|}{5.27} & \multicolumn{1}{c|}{0.06}  & \multicolumn{1}{c|}{4.26} & \multicolumn{1}{c|}{5.56} & \multicolumn{1}{c|}{0.02}  & \multicolumn{1}{c|}{4.31} & \multicolumn{1}{c|}{5.30} & \multicolumn{1}{c|}{0.03}  & \multicolumn{1}{c|}{4.18} & \multicolumn{1}{c|}{5.37} & 0.04  & \multicolumn{1}{c|}{4.42}                 & \multicolumn{1}{c|}{5.46}                  & 0.08               \\ 
                               & 14                                                                                        & \multicolumn{1}{c|}{4.33} & \multicolumn{1}{c|}{5.42} & \multicolumn{1}{c|}{0.03}  & \multicolumn{1}{c|}{4.28} & \multicolumn{1}{c|}{5.78} & \multicolumn{1}{c|}{0.02}  & \multicolumn{1}{c|}{4.26} & \multicolumn{1}{c|}{5.31} & \multicolumn{1}{c|}{-0.02} & \multicolumn{1}{c|}{4.29} & \multicolumn{1}{c|}{5.40} & -0.07 & \multicolumn{1}{c|}{4.98}                 & \multicolumn{1}{c|}{5.95}                  & 0.12               \\ 
                               & 16                                                                                        & \multicolumn{1}{c|}{4.50} & \multicolumn{1}{c|}{5.56} & \multicolumn{1}{c|}{0.05}  & \multicolumn{1}{c|}{4.58} & \multicolumn{1}{c|}{5.79} & \multicolumn{1}{c|}{0.01}  & \multicolumn{1}{c|}{4.62} & \multicolumn{1}{c|}{5.76} & \multicolumn{1}{c|}{-0.03} & \multicolumn{1}{c|}{4.62} & \multicolumn{1}{c|}{5.82} & -0.06 & \multicolumn{1}{c|}{5.26}                 & \multicolumn{1}{c|}{6.20}                  & 0.07               \\  
                               & 60                                                                                        & \multicolumn{1}{c|}{5.13} & \multicolumn{1}{c|}{6.35} & \multicolumn{1}{c|}{-0.07} & \multicolumn{1}{c|}{5.10} & \multicolumn{1}{c|}{6.22} & \multicolumn{1}{c|}{0.06}  & \multicolumn{1}{c|}{5.21} & \multicolumn{1}{c|}{6.46} & \multicolumn{1}{c|}{-0.08} & \multicolumn{1}{c|}{5.25} & \multicolumn{1}{c|}{6.50} & -0.17 & \multicolumn{1}{c|}{5.27}                 & \multicolumn{1}{c|}{6.48}                  & 0.18               \\ \hline
\end{tabular}
\end{threeparttable}
\end{adjustbox}
\end{table*}

\begin{table*}[!h]
\centering
\caption{Performance comparison of our proposed method against its Eulerian-only and Lagrangian-only variants, as well as existing baseline approaches for HR estimation.}
\label{tab:my-table-hr}
\begin{adjustbox}{width=0.8\linewidth}
\begin{threeparttable}
\begin{tabular}{|c|c|ccc|ccc|ccc|}
\hline
\multirow{3}{*}{\textbf{Modality}} & \multirow{3}{*}{\textbf{Methods}} & \multicolumn{3}{c|}{\textbf{RR-rPPG} \& \textbf{COHFACE}} & \multicolumn{3}{c|}{\textbf{RR-rPPG}} & \multicolumn{3}{c|}{\textbf{COHFACE}} \\ \cline{3-11}
& & \textbf{MAE} & \textbf{RMSE} & \textbf{r} & \textbf{MAE} & \textbf{RMSE} & \textbf{r} & \textbf{MAE} & \textbf{RMSE} & \textbf{r} \\ \hline

\multirow{5}{*}{Eulerian}
& HR-Green & 2.28 & 2.48 & 0.74 & 2.30 & 2.62 & 0.71 & 2.45 & 2.50 & 0.72 \\
& HR-Red & 2.39 & 2.54 & 0.69 & 2.50 & 2.78 & 0.62 & 2.63 & 2.88 & 0.67 \\
& HR-Blue & 2.48 & 2.67 & 0.62 & 3.54 & 3.76 & 0.51 & 2.64 & 2.90 & 0.65 \\
& HR-Chrom & 2.31 & 2.49 & 0.72 & 2.37 & 2.72 & 0.68 & 2.55 & 2.77 & 0.69 \\
& HR-SI (ours) & \textbf{1.76} & \textbf{1.84} & \textbf{0.79} & \textbf{1.99} & \textbf{2.11} & \textbf{0.78} & \textbf{1.75} & \textbf{2.01} & \textbf{0.76} \\ \hline

\multirow{2}{*}{Lagrangian}
& HR-WD & 2.01 & 2.11 & 0.67 & 2.00 & 2.14 & 0.54 & 2.11 & 2.23 & 0.68 \\
& HR-D (ours) & \textbf{1.47} & \textbf{1.58} & \textbf{0.84} & \textbf{1.75} & \textbf{1.82} & \textbf{0.81} & \textbf{1.57} & \textbf{1.58} & \textbf{0.80} \\ \hline

Multimodal
& Proposed & \textbf{1.01} & \textbf{1.27} & \textbf{0.90} & \textbf{1.30} & \textbf{1.65} & \textbf{0.86} & \textbf{1.50} & \textbf{1.52} & \textbf{0.83} \\ \hline
\end{tabular}
\begin{tablenotes}
\item[Note] All above results are reported for a video clip length of 12 seconds.
\end{tablenotes}
\end{threeparttable}
\end{adjustbox}
\end{table*}

\subsection{Discussions}
\begin{figure}
    \centering
    \includegraphics[width=0.9\linewidth]{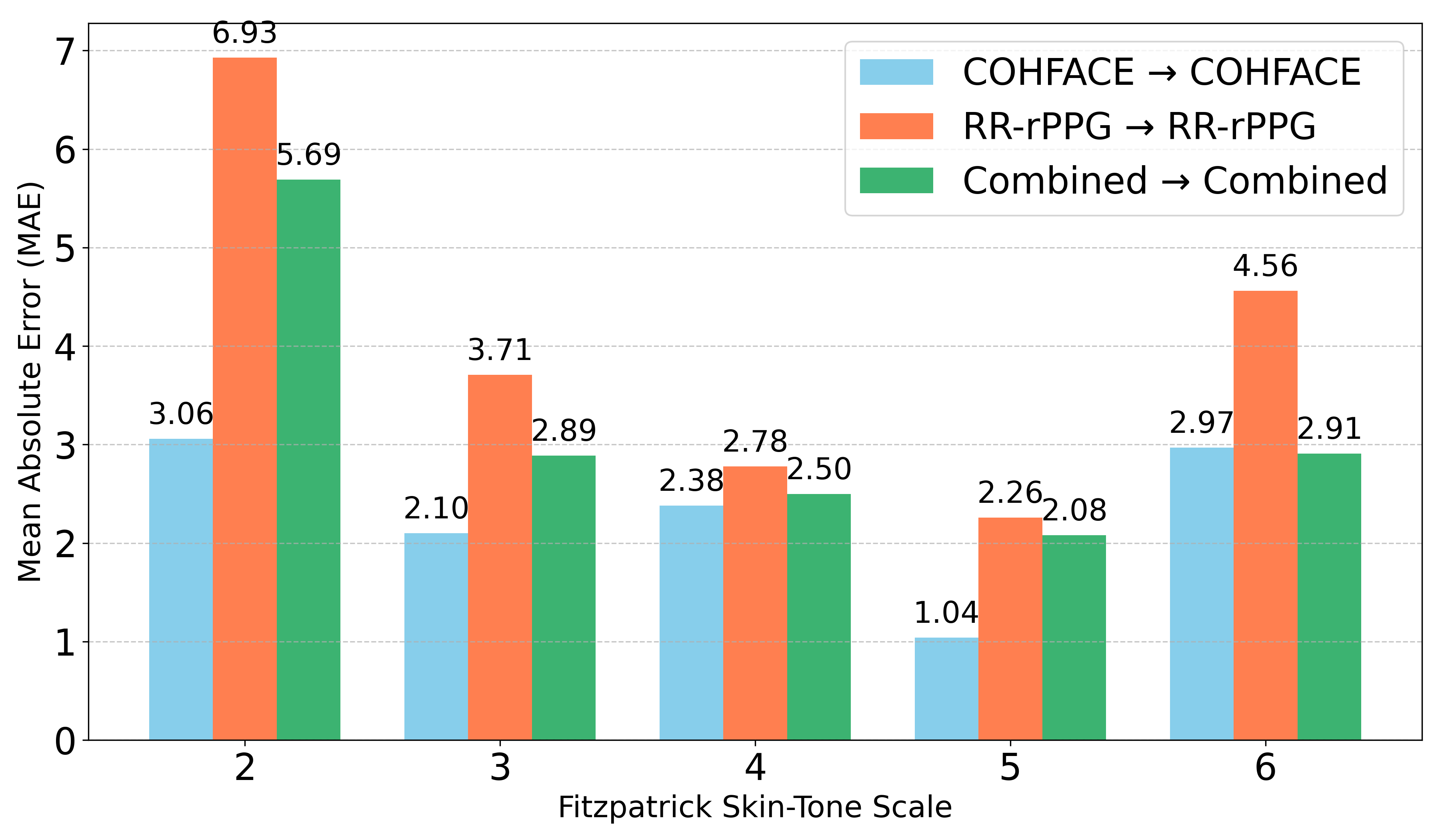}
    \caption{MAE across Fitzpatrick skin-tone scales for different training and testing dataset combinations (``X → Y'' denotes that the model was trained on dataset X and tested on dataset Y).}
    \label{fig:mae-skin-tone}
\end{figure}

To assess the robustness of \textit{ELITE-RR} under varying skin-tones, we analyzed performance across Fitzpatrick scales using the RR-rPPG and COHFACE datasets (Figure \ref{fig:mae-skin-tone}). It consistently yields lower MAE values across most skin tones, particularly on scales 3–5, indicating its generalizability. Notably, even for darker skin tones (scale 6), where rPPG estimation is typically challenging, the method trained on combined datasets maintains comparable performance (MAE: 2.91), outperforming single-dataset training settings, highlighting the effectiveness of our skin-tone-aware signal formulation. However, we observe relatively higher errors for extremely light (scale 2) and dark (scale 6) tones in certain training conditions, primarily due to the limited number of samples in these groups, as shown in Figure \ref{fig:fitzpatrick}. This suggests that performance can further benefit from a more balanced and diverse dataset.

To evaluate the broader applicability of our proposed skin-tone–aware method, we extended it to the task of HR estimation. Table \ref{tab:my-table-hr} presents a comparative analysis of HR estimation performance across multiple baseline methods and our proposed models, under both Eulerian and Lagrangian representations.
Our Eulerian variant (HR-SI) consistently outperforms standard rPPG methods such as HR-Green, HR-Red, HR-Blue and HR-Chrom in all datasets, demonstrating a lower MAE and RMSE; and higher $r$. Similarly, in the Lagrangian representation, our denoising-based model (HR-D) shows notable improvements over the baseline HR-WD method.
Most importantly, the dual-representation variant of our method achieves the best overall performance, with an MAE of 1.01 and $r = 0.90$ on the combined RR-rPPG \& COHFACE dataset. These results validate the effectiveness and generalizability of our proposed method beyond RR estimation, affirming its capability in robust and enhanced remote HR estimation.

\section{Conclusion}
\label{sec:conclusion}
We presented a novel dual-representation method, \textit{ELITE-RR}, for remote RR estimation that leverages both Eulerian and Lagrangian representations. To enhance the reliability of rPPG signals extracted through Eulerian representation, we have introduced a skin-tone–aware projection network that extracts the most informative physiological signal. For the Lagrangian representation, a dedicated denoising network has been designed to suppress non-respiratory artifacts and refine motion-based signals. To facilitate effective knowledge transfer between representation, we have proposed a phase-independent contrastive loss. Alongside the methodology, we contributed a new RR-centric dataset, RR-rPPG, and conducted extensive evaluations on both RR-rPPG and the publicly available COHFACE dataset. \textit{ELITE-RR} consistently outperforms existing methods across multiple settings. Future work includes expanding the dataset to diverse, real-world scenarios and extending our method to other physiological parameters.

\section*{CRediT authorship contribution statement}
\noindent
\textbf{Trishna Saikia:} Conceptualization, Methodology, Data curation, Formal analysis, Investigation, Validation, Visualization, Writing - original draft. 
\textbf{Anup Kumar Gupta:} Conceptualization, Methodology, Data curation, Investigation, Validation, Writing - review \& editing. 
\textbf{Puneet Gupta:} Conceptualization, Methodology, Supervision, Funding acquisition, Project administration, Writing - review \& editing. 
\textbf{Pasi Liljeberg:} Supervision, Writing - review \& editing.

\section*{Declaration} 
This study involved human participants. Approval for all ethical, experimental procedures and protocols was granted by the Institutional Human Ethics Committee (IHEC approval no: BSBE/IITI/IHEC-10/2024-3), Indian Institute of Technology Indore, India.\\

\noindent
\textbf{Declaration of generative AI and AI-assisted technologies in the manuscript preparation process}: During the preparation of this work, the authors used the assistance of LLM in order to generate and edit the graphical abstract. After using the tool, the authors reviewed and edited the content as needed and take full responsibility for the content of the published article.

\bibliography{main}
\bibliographystyle{unsrtnat}

\end{document}